\documentclass[10pt,journal,twocolumn,twoside]{IEEEtran} 

\usepackage{graphicx}
\usepackage{epstopdf}
\usepackage{float}
\usepackage{algorithmic}
\usepackage{array}
\usepackage{amsmath}
\usepackage{amssymb}
\usepackage{mdwmath}
\usepackage{mdwtab}
\usepackage{eqparbox}
\usepackage{stfloats}
\usepackage{fixltx2e}
\usepackage{cases} 

\usepackage{flushend}

\usepackage{xcolor}
\usepackage{makecell}

\usepackage[boxed,ruled,commentsnumbered]{algorithm2e}
\usepackage{url}
\usepackage{cite}
\usepackage{upgreek}

\ifCLASSOPTIONcompsoc
\usepackage[caption=false,font=normalsize,labelfont=sf,textfont=sf]{subfig}
\else
\usepackage[caption=false,font=footnotesize]{subfig}
\fi
\allowdisplaybreaks[4]

\newtheorem{remark}{Remark}

\makeatletter
\renewcommand*{\@opargbegintheorem}[3]{\trivlist
      \item[\hskip \labelsep{\bfseries #1\ #2}] \textbf{(#3):}\ }
\makeatother

\begin{document}

\title
{Reconfigurable Intelligent Surface Enabled Joint Backscattering and Communication}
\author{Jinqiu Zhao, Jia~Ye,~\IEEEmembership{Member,~IEEE,} Shuaishuai Guo,~\IEEEmembership{Senior Member,~IEEE,} Zhiquan Bai, \IEEEmembership{Senior Member, IEEE}, Di Zhou, and Abeer Mohamed
    
\thanks{Jinqiu Zhao, Zhiquan Bai,  Di Zhou and Abeer Mohamed are with Shandong Provincial Key Lab. of Wireless Communication Technologies, School of Information Science and Engineering, Shandong University, Qingdao 266237, China (e-mail: 202020373@mail.sdu.edu.cn, zqbai@sdu.edu.cn,  emailofzhoudi@163.com, abeermohamed@mail.sdu.edu.cn).}
\thanks{Jia Ye is with the school of Electrical Engineering, Chongqing University, Chongqing, 400044, China (e-mail: yejiaft@163.com).}
\thanks{Shuaishuai Guo is with Shandong Provincial Key Lab. of Wireless Communication Technologies, School of Control Science and Engineering, Shandong University, Jinan, 250061, China (e-mail: shuaishuai\textunderscore guo@sdu.edu.cn).}

\thanks{Accepted by IEEE Transactions on Vehicular Technology}
   }
\maketitle


\begin{abstract} 
Reconfigurable intelligent surface (RIS) as an essential topic in the sixth-generation (6G) communications aims to enhance communication performance or mitigate undesired transmission. However, the controllability of each reflecting element on RIS also enables it to act as a passive backscatter device (BD) and transmit its information to reader devices. In this paper, we propose a RIS-enabled joint backscattering and communication (JBAC) system, where the backscatter communication coexists with the primary communication and occupies no extra spectrum.
Specifically, the RIS modifies its reflecting pattern to act as a passive BD and reflect its own information back to the base station (BS) in the backscatter communication, while helping the primary communication from the BS to the users simultaneously. We further present an iterative active beamforming and reflecting pattern design to maximize the user average transmission rate of the primary communication and the goodput of the backscatter communication by solving the formulated multi-objective optimization problem (MOOP). Numerical results fully uncover the impacts of the number of reflecting elements and the reflecting patterns on the system performance, and demonstrate the effectiveness of the proposed scheme. Important practical implementation remarks have also been discussed.


\end{abstract}


\section{Introduction} 
\IEEEPARstart{T}{he} past decade has witnessed explosive growth in wireless-connected devices and intelligent appliances. The prosperity of humankind has never been so reliant on wireless connectivity. Providing high performance and quality of service for all terminals in a green manner is one of the key challenges of the 6th generation (6G) communications \cite{Basar2020,Li2022,Basharat2022}, which has aroused enthusiasm in the wireless society to explore low power consumption communication techniques. A promising technology for achieving the energy-saving target is backscatter communication, where the backscatter device (BD) delivers its messages by reflecting the signals from an external emitter to the backscatter reader \cite{niu2019overview,Zhang2020a,Liang2022}. The conveyed messages are encoded by the reflection coefficients at the BD. Under this setup, the BD does not need the active radio frequency components by utilizing the external signal carriers, which greatly cuts off the system energy consumption. To unleash the potential of backscatter communication, various research works have been conducted to improve the communication range \cite{belo2019iq}, transmission rate \cite{daskalakis2018four,Xu2023}, bit error rate (BER) \cite{ElMossallamy2019}, and energy efficiency (EE) \cite{Xu2021a,Xu2021} of the backscatter communication system.

In backscatter communication, the signals may experience double-fading effect introduced in the forward link from the emitter to the BD and the backward link from the BD to the reader, which seriously limits the system performance \cite{Liang2022}. The recently proposed cutting-edge technology, reconfigurable intelligent surface (RIS), has rich potential in boosting the received signal quality, owing to its prominent capability of reconfiguring the propagation environments \cite{Wu2020,xurisa}. RIS is an artificial planar surface composed of many low-cost and small-size passive reflecting elements, which can be realized by newly developed advanced microelectrical-mechanical systems and metamaterials. Another similar technology is holographic multiple-input multiple-output surfaces (HMIMOS), which is composed of sub-wavelength patch antennas to realize programmable wireless environments \cite{Huang2020}. Each reflecting element of RIS is able to reflect the incident signals with controllable amplitude and phase-shift, thereby achieving the enhancement of the directional signal and the reconfiguration of the communication environments. Through the appropriate design of RIS phase shifts, the performance of RIS-enabled system can be improved significantly \cite{xurisc,xurisb}. Besides the traditional design methods, deep reinforcement learning (DRL) is taken to design the phase shifts of RIS by learning from the dynamic environments \cite{Guo2023,Peng2022}. Moreover, RIS can also be used in the joint communication and sensing of the indoor multi-user uplink communication \cite{Tong2021}. The properties of RIS have further motivated the research of RIS based backscatter communication \cite{Liu2022,Fara2022,Asiedu2022}.

Given the widespread deployment of RIS on objects, it can assist the forward link or the backward link of the backscatter system to enhance the desired performance. For example, by deploying a RIS between multiple BDs and the reader, the phase shifts at the RIS and the power reflection coefficients at the BDs were jointly optimized in \cite{zuo2021reconfigurable} to enhance backscatter links. In \cite{chen2021joint}, a RIS was deployed between an emitter and multiple BDs for the RIS-enhanced symbiotic system, which achieves lower transmission power than the system without RIS. Moreover, a RIS was considered in \cite{jia2020intelligent,jia2021irs} to reflect the signals from both the emitter and the BD, which introduces a highly non-trivial design problem. Multiple BDs were considered in the following work \cite{jia2021intelligent}, the transmit power minimization problem with the constraints of BDs' signal-to-noise ratio (SNR) requirements was studied. The authors in \cite{chen2021performance} analyzed the BER of the RIS-assisted backscatter system, where the system BER tends to decrease when RIS is employed to enhance the direct link between the emitter and the reader, and increase if RIS is deployed to enhance the forward link between the emitter and the BD. Moreover, three RISs were employed in \cite{nemati2020short} to assist the direct link, the forward link, and the backward link, respectively, and achieved a better system BER performance. The above studies verify the importance of RIS in reconfiguring the forward link and the backward link, and enhancing the system BER performance. 

In fact, both the BD and the RIS can work in full-duplex passive reflection mode for a hardware-cost-saving and energy-saving manner without active transmission chains. These circumstances enable RIS to work not only as a helper to assist the backscatter links, but also as a BD to develop the backscatter system. However, there has limited progress in the research on RIS operating as a BD for information transmission. The authors in \cite{zhou2020cooperative, hu2020reconfigurable, zhang2021reconfigurable} proposed the design of the reflecting coefficients at the RIS to realize the novel RIS-enabled backscatter system by denoting the reflection phase-shift matrix and the transmitted symbol at the RIS as $\pmb{\Phi}$ and $c = \{-1,1\}$, respectively. Thus, the information carried by RIS can be delivered to the backscatter reader by reflecting the incident signals transmitted from the emitter with the designed reflecting coefficient $c\pmb{\Phi}$. The transmitted symbol at the RIS can be modulated by the binary phase-shift keying (BPSK) \cite{zhou2020cooperative, zhang2021reconfigurable}, or follows the standard normal distribution \cite{hu2020reconfigurable}. Additionally, the authors in \cite{park2020intelligent} designed a RIS-aided backscatter system, where the phase shifts of the RIS can be optimized and updated by adding it with BPSK information bits, 0 or $\pi$, to convey its own information.

The multiplication or addition operations on the phase shifts may impair the achieved performance gain provided by the RIS. Moreover, the predefined symbol format is irrelevant to the channel state information (CSI) and the system settings, which cannot ensure its superiority in all communication scenarios. Based on these facts, developing multiple reflecting patterns and letting the unique index of each reflecting pattern as the transmitted signal at the RIS will be a good option. This idea has already been implemented in \cite{Guo2020,Ye2022} to enable RIS to carry additional information when assisting the transmission between the conventional transmitter and receiver. However, it has not been considered in the backscatter communication system. The backscatter communication is generally designed to coexist with the primary communication from the  base station (BS) to the user to share the transmitter and spectrum resources in most practical scenarios. To the best of our knowledge, the functionalities of RIS in the joint backscattering and communication system have not been well-investigated. 

To bridge the gap in the RIS design for both enabling the backscatter communication and enhancing the primary communication, we propose a RIS-enabled joint backscattering and communication (JBAC) system. In particular, the RIS employs multiple reflecting patterns to encode its own information, which operates as the modulator of the BD in the backscatter communication. This method provides a practical way to achieve the low-cost communication between RIS and BS, and can be used in Internet-of-Things networks.
Meanwhile, the incident signals transmitted from the BS are reflected by RIS with the selected reflecting pattern to the served users in primary communication. As the exact achievable rate is hard to derive in backscatter communication, we introduce another BER-depending performance metric, named goodput\cite{Wang2008}, to evaluate the successfully delivered information bits from the RIS to the BS. The reflecting patterns at the RIS and the transmit beamforming at the BS are jointly designed to maximize the user average transmission rate of the primary communication and the goodput of the backscatter communication system. The main contributions are listed as follows.


\begin{itemize}
\item 
 A novel RIS-enabled JBAC system is proposed, where the BS can work as the emitter and the reader at the same time, while the RIS assists the primary communication and operates as the BD to convey its information to the BS. The information from the BD to the BS is carried by the index of the reflecting pattern selected from the reflecting candidate set, which needs to be recovered at the BS. Thus, the backscatter communication system can coexist with the primary communication system by sharing the same transmission and channel resources. We analyze the user average transmission rate of the primary communication and the goodput as well as its BER of the backscatter communication to evaluate the performance of the proposed system.
  \item Based on the analyzed performance metrics, a multi-objective optimization problem (MOOP) is formulated to maximize the user average transmission rate of the primary communication and the goodput of the backscatter communication. To facilitate the solution of the MOOP, we further reformulate the MOOP through the weighted sum scalarization method. By adopting the alternating algorithm, an efficient active beamforming and reflecting pattern design scheme is presented to optimize the proposed system. 
 \item Numerical results illustrate that the weighting parameter should be carefully chosen to strike a favorable performance tradeoff between the primary communication and the backscatter communication. The impacts of the number of reflecting elements and the reflecting patterns on the system performance have been presented in detail. The superiorities of the proposed system design and alternating algorithm are verified through the comparison with the existing RIS based information-carrying systems. The impacts of the imperfect CSI and discrete phase shifts on the proposed JBAC system are also discussed. We also find that the satisfied performance can be achieved by setting a reasonable number of discrete phase shifts.
\end{itemize}

The remainder of the paper is organized as follows. Section \ref{JBAC} describes the proposed RIS-enabled JBAC system and provides the corresponding performance metrics. In Section \ref{H}, a MOOP of the proposed system is formulated to enhance the goodput of the backscatter communication and the average transmission rate of the primary communication. A feasible solution based on the alternating algorithm to optimize the active beamforming at the BS and the reflecting patterns at the RIS is proposed. Numerical results are demonstrated in Section \ref{SR}, and conclusions are drawn in Section \ref{CON}.

\emph{Notations}: In this paper, $\mathbb{C}$ represents the complex domain, $\mathbf{(\cdot)}^T$ and $\mathbf{(\cdot)}^H$ are the transpose and conjugate transpose, respectively. $\mathcal{CN}\left(0, {\mu}^2\right)$ stands for the circularly symmetric complex Gaussian distribution with mean $0$ and covariance ${\mu}^2$. $\mathbb{E}[\cdot]$ is the expectation operation. $\operatorname{diag}({\mathbf{X})}$ is  a vector whose entries come from the diagonal entries of matrix $\mathbf{X}$. $\operatorname{diag}\{\mathbf{x}\}$ returns a diagonal matrix composed of the elements of vector $\mathbf{x}$. $|.|$ is the absolute value. $||\cdot||$, $||\cdot||_{\infty}$, $||\cdot||_{p}$ are the $l_2$ norm, $l_{\infty}$ norm, and $l_p$ norm, respectively. $\binom{\cdot}{\cdot}$ is a binomial
coefficient. $\mathbf{I}_m$ represents the $m \times 1$ all-one vector.

\section{RIS-Enabled JBAC System}\label{JBAC}
\subsection{System Model}

\begin{figure}
  \centering
  \includegraphics[width=0.48\textwidth]{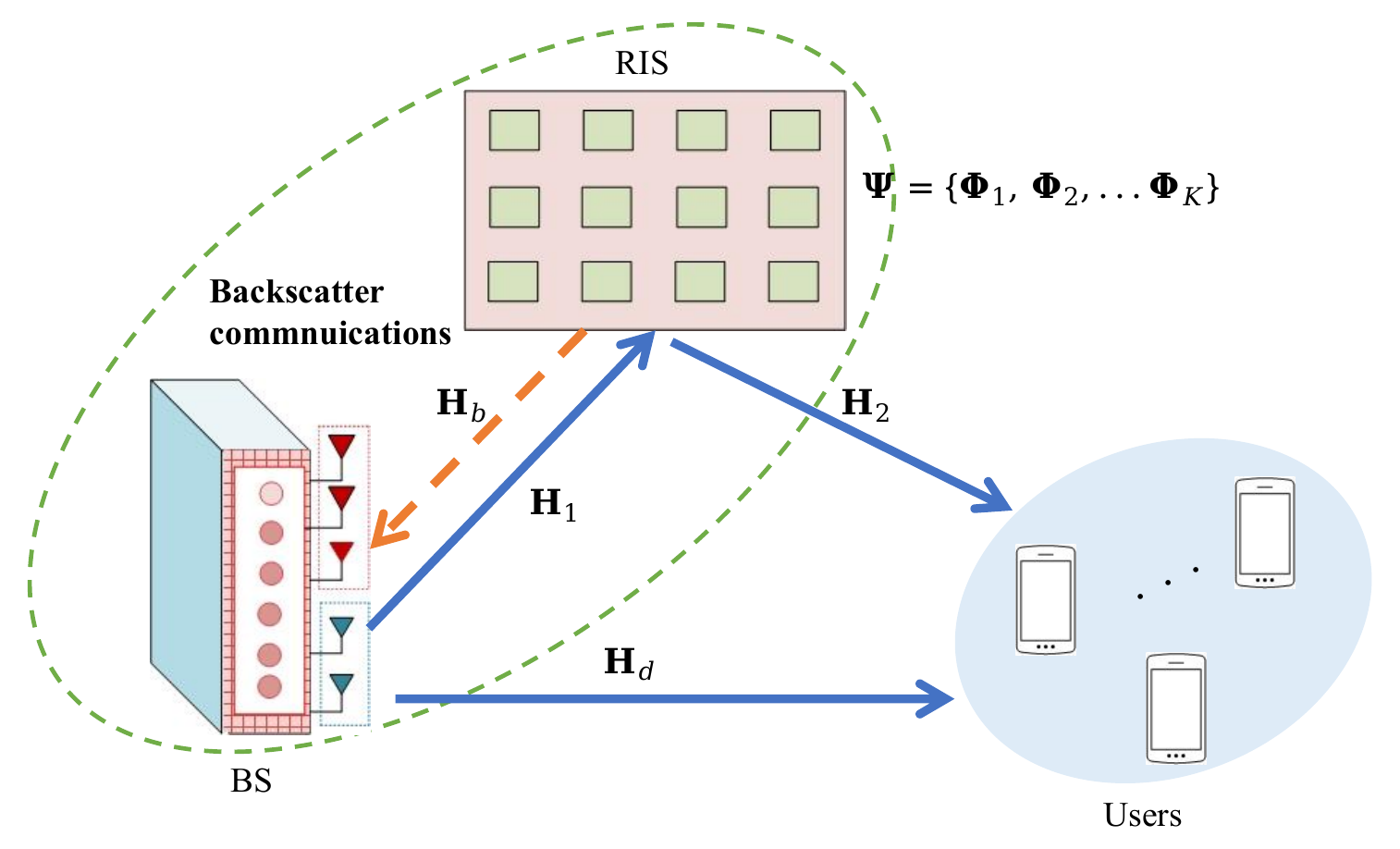}
  \caption{RIS-enabled JBAC system.} 
  \label{fig1}
\end{figure}

As shown in Fig. 1, we consider a RIS-enabled JBAC system, where a BS communicates with multiple users with the assistance of RIS and acts as a reader to decode the information from the RIS in the backscatter communication. The BS multicasts the signal to all the users in the primary communication. In this work, the backscatter communication system coexists with the RIS-assisted primary communication system. The backscatter communication consumes no additional channel resources by utilizing the spectrum of the primary communication. 

The proposed RIS-enabled JBAC system requires that BS has the ability to transmit and receive signals simultaneously. Therefore, the BS is assumed to work in the full-duplex mode with $N_{t}$ transmitting antennas and $N_{r}$ receiving antennas, which can simultaneously transmit and receive signals without interference. It is noteworthy that the full-duplex assumption is reasonable as the BS with integrated sensing and communication (ISAC) capabilities is a new trend and it requires the full-duplex hardware structure. There are $M$ user terminals equipped with single receiving antenna and located within the coverage of BS and RIS. The RIS is composed of $N$ reflecting elements, which can be embedded on the sensor to transmit the sensed information or the power-limited terminal, such as the aerial platforms to convey the channel or environment conditions to the BS. 

The channels between any two devices in the proposed system are considered to follow the block flat-fading, whose channel coefficients remain unchanged during a block but may vary from one block to another. The length of the block is denoted as $T_c$. As shown in Fig. 1, the channel matrices between the BS and the RIS, between the RIS and the users, between the BS and the users, between the RIS and the BS are denoted by $\mathbf{H}_{1} \in \mathbb{C}^{ N\times N_{t}}$, $\mathbf{H}_{2} = \left[\mathbf{h}_{2,1}, \mathbf{h}_{2,2}, \cdots, \mathbf{h}_{2,M}\right] \in \mathbb{C}^{N\times M}$, $\mathbf{H}_{d} = \left[\mathbf{h}_{d,1}, \mathbf{h}_{d,2}, \cdots, \mathbf{h}_{d,M}\right] \in \mathbb{C}^{N_{
t}\times M}$, $\mathbf{H}_{b} \in \mathbb{C}^{N_{r} \times N}$, respectively, where $\mathbf{h}_{2,m} \in \mathbb{C}^{N\times 1}$ and $\mathbf{h}_{d,m} \in \mathbb{C}^{N_t\times 1}$ respectively represent the channel between the RIS and the $m$-th user, and between the BS and the $m$-th user. $\mathbf{H}_{1}$ and $\mathbf{H}_{b}$ are different. 
Each user sends their own pilots sequentially, which are reflected from the RIS to the BS. Once the CSI of the uplink is obtained at the BS, the CSI of the downlink can be obtained by the channel reciprocity \cite{DaiL2021}. The pilot signals from the BS are reflected by the RIS back to the BS, and the CSI of backscatter communication is obtained with the technique in \cite{He2020}. For the training overhead of the system is smaller than the block length, we ignore the time of channel estimation. For simplicity, we assumed that all channels are perfectly known in this work. Since the channel estimation errors are inevitable sometimes,  we also investigate the proposed system performance under imperfect CSI in Section IV.  


\subsection{Signal Processing}
We assume that each flat-fading block covers $L$ ($L \gg 1$) symbol periods. The $l$-th transmitted symbol from the BS to the users in the primary communication system follows the Gaussian distribution denoted as $s(l) \sim \mathcal{CN}(0,1)$. Denoting the duration of each symbol as $T_s$, we have $T_c \ge LT_s$. For the backscatter communication in the RIS-enabled JBAC system, the information delivered by the RIS is modulated by the index of the reflecting pattern equiprobably selected from the reflecting candidate set $\pmb{\Psi} = \{\pmb{\Phi}_{1}, \pmb{\Phi}_{2}, \cdots, \pmb{\Phi}_{K}\}$ with size $K$, where $\pmb{\Phi}_{k}$ is the diagonal reflecting matrix with index $k$, whose elements satisfying $\left(\pmb{\Phi}_{k}\right)_{n, n} \in\{0\} \cup\left\{e^{j \theta}, \theta \in [0, 2\pi) \right\}$, $\forall n \in\{1,2, \ldots, N\}$. The reflecting candidate set $\pmb{\Psi}$ is assumed to remain unchanged during the communication block, and a reflecting pattern is re-selected over each symbol duration within a communication block. It means one of the reflecting patterns will be selected from the reflecting candidate set and applied on the RIS to reflect $s(l)$ to users during each symbol duration. Meanwhile, the BS can also receive the signal reflected from the RIS to recover the index of the selected reflecting pattern, which carries the information from the RIS in the backscatter communication system. 

\subsection{Performance Metrics}
\subsubsection{Primary Communication System}
Let $\mathbf{w}\in \mathbb{C}^{ N_{t}\times 1}$ denote the active beamforming vector at the BS and $\pmb{\Theta} \in \pmb{\Psi}$ be the selected reflecting pattern matrix in the transmission of the $l$-th multicast symbol $s(l)$. The received signal at the $m$-th user,  ${y}_{m}(l)$, $l=1,2 \ldots, L$, is given as
\begin{equation}
y_m(l)=\left(\mathbf{h}_{2,m}^H \pmb{\Theta} \mathbf{H}_{1}+\mathbf{h}_{d,m}^H\right)\mathbf{w} s(l)+{n}_{m}(l),
\end{equation}
where ${n}_{m}(l)$ is the additive white Gaussian noise (AWGN) at the $m$-th user within the $l$-th symbol duration that follows $\mathcal{CN}\left(0, \sigma_{m}^2\right)$.
The received SNR at the $m$-th user can be calculated as

\begin{equation}
\gamma_{m}(\pmb{\Theta})=\frac{|(\mathbf{h}_{2,m}^H \pmb{\Theta} \mathbf{H}_{1}+\mathbf{h}_{d,m}^H)\mathbf{w}|^{2}}{\sigma_{m}^2}.
\end{equation}

The achievable rate of the $m$-th user can be expressed as
\begin{equation}
R_{m}=\frac{1}{T_{s}} \mathbb{E}_{\pmb{\Theta}}\left[\log _{2} \left(1+\gamma_{m}(\pmb{\Theta})\right)\right].
\end{equation}
We assume that the reflecting patterns in $\pmb{\Psi}$ are uniformly activated. Thus, the expectation in (3) can be removed and (3) is rewritten as       
\begin{equation}
R_{m}=\frac{1}{T_{s}K} \sum_{k=1}^K\log _{2} \left(1+\gamma_{m}(\pmb{\Phi}_{k})\right).
\end{equation}
 
Accordingly, the user average transmission rate of the primary communication system can be calculated as
\begin{equation}
R_u = \frac{1}{M} \sum_{m=1}^M R_{m} = \frac{1}{T_{s}MK} \sum_{m=1}^M\sum_{k=1}^K\log _{2} \left(1+\gamma_{m}(\pmb{\Phi}_k)\right).
\end{equation}

\subsubsection{Backscatter Communication System}
The received signal $\mathbf{y}_{b}(l)\in \mathbb{C}^{ N_{r}\times 1}$ at the BS in the backscatter communication system for the transmitted symbol $s(l)$ and the reflecting pattern $\pmb{\Theta} \in \pmb{\Psi}$ at the RIS can be expressed as

\begin{equation}
\begin{split}
\mathbf{y}_{b}(l)&=\mathbf{H}_{b} \pmb{\Theta} \mathbf{H}_{1} \mathbf{w}{s}(l)+\mathbf{n}_{b}(l)\\
& = \bar{\mathbf{H}}\pmb{\uptheta}+\mathbf{n}_{b}(l),	
\end{split}
\end{equation}
where $\mathbf{n}_{b}(l)$ follows $\mathcal{CN}\left(0, \mathbf{I}_{N_{r}} N_{0}\right)$ and represents the AWGN vector at the BS, $\bar{\mathbf{H}} = \mathbf{H}_{b}{\operatorname{diag}}\{\mathbf{H}_{1} \mathbf{w}{s}(l)\}$, and $\pmb{\uptheta}=\operatorname{diag}(\pmb{\Theta})$.

When the BS knows $\bar{\mathbf{H}}$, the achievable rate of the backscatter communication system can be derived based on the mutual information between $\pmb{\uptheta}$ and $\mathbf{y}_b(l)$ as
\begin{equation}\label{MI}
\begin{split}
&R_{b} = \mathcal{I}(\pmb{\uptheta} ; \mathbf{y}_b(l) \mid \bar{\mathbf{H}}) =\log _2K-\frac{1}{K} \\
&\times \sum_{k_1=1}^{K} \mathbb{E}_{\mathbf{n}_b(l)}\left\{\log _2 \sum_{k_2=1}^{K} e^{-\frac{\left\|\bar{\mathbf{H}}\left(\pmb{\upphi}_{k_1}-\pmb{\upphi}_{k_2}\right)+\mathbf{n}_b(l)\right\|^2-\|\mathbf{n}_b(l)\|^2}{N_0}}\right\}, 
\end{split}
\end{equation}
with $\pmb{\upphi}_k = \operatorname{diag}(\pmb{\Phi}_k)$ for $k \in \{1,2, \cdots, K\}$.

As \eqref{MI} reveals, the mutual information of the backscatter communication system decreases with $N_0$, which is upper bounded by $\log _2K$ when $N_0$ goes to zero. Therefore, the maximum achievable rate of the backscatter communication enabled by RIS is $\frac{1}{T_s}\log _{2}K$ bits/s, which depends on the size of the generated reflecting candidate set.

BER is an important performance in the considered backscatter communication system and related to another important performance named goodput\cite{Wang2008}. In the following, we study the goodput of the proposed backscatter communication system. 

With $\mathbf{H}_b$ and $\mathbf{H}_1$ being known by the BS, the maximum likelihood (ML) detection is employed, and the information delivered by the RIS can be recovered from
\begin{equation}
\tilde{k}=\arg \min _{k}\left\|\mathbf{y}_b(l)-\mathbf{H}_b \boldsymbol{\Phi}_k \mathbf{H}_1 \mathbf{w} {s}(l)\right\|^2
\end{equation}

As the closed-form BER of the backscatter system is hard to derive, we can calculate its upper bound as a replacement and it is computed as
\begin{equation}\label{pe}
P_{e} \leq \bar{P}_{e}=\frac{1}{K\log _{2} K} \sum_{k=1}^{K} \sum_{\hat{k}=1, \hat{k} \neq k}^{K} D_{\mathrm{HD}}\left({k}, {\hat{k}}\right) P_{\mathrm{e}}(k \rightarrow \hat{k}),
\end{equation}
where $D_{\mathrm{HD}}\left({k}, {\hat{k}}\right)$ represents the Hamming distance between the coded reflecting pattern indices ${k}$ and ${\hat{k}}$, and $P_{\mathrm{e}}(k \rightarrow \hat{k})$ denotes the corresponding pairwise error probability and can be expressed by
\begin{equation}
P_{e}(k \rightarrow \hat{k})=\mathrm{Q}\left(\sqrt{\frac{D_{\mathrm{ED}}(k \rightarrow \hat{k})^{2}}{2N_{0}}}\right),
\end{equation}
where $Q\left(\cdot\right)$ is the $Q$-function, and $D_{\mathrm{ED}}(k \rightarrow \hat{k})$ means the Euclidean distance between the two received noise-free signal vectors calculated as
\begin{equation}\label{DED}
\begin{split}
D_{\mathrm{ED}}(k \rightarrow \hat{k})&=\left\|\mathbf{H}_{b} \pmb{\Phi}_{k} \mathbf{H}_{1}\mathbf{w}s(l)- \mathbf{H}_{b} \pmb{\Phi}_{\hat{k}} \mathbf{H}_{1}\mathbf{w}s(l)\right\|\\
& = \left\|\bar{\mathbf{H}}\pmb{\upphi}_{k}- \bar{\mathbf{H}}\pmb{\upphi}_{\hat{k}}\right\|\\
& = \left\|\bar{\mathbf{H}}\left(\pmb{\upphi}_{k}- \pmb{\upphi}_{\hat{k}}\right)\right\|.
\end{split}
\end{equation}

As the mutual information is not in a closed form and has an upper bound, we may focus on the goodput performance of the backscatter system decided by the system BER. The goodput can be defined as the number of successfully delivered information bits from the RIS to the BS per second and written as
\begin{equation}\label{GP}
\begin{split}
G_{b}=\frac{\log _{2} K\left(1-P_{e}\right)}{T_s}.
\end{split}
\end{equation}
It is noteworthy that we are interested in the average BER of the backscatter communication for $s(l)$ is a random variable. By substituting the expectation of BER upper bound $\mathbb{E}_{s(l)}[\bar{P}_{e}]$ into \eqref{GP}, the lower bound of the goodput $\bar{G}_{b}$ can be derived accordingly. 

\section{MOOP Optimization}\label{H}
In this section, based on the performance metrics analyzed above, we first formulate a MOOP to enhance the primary communication and the backscatter communication of the proposed RIS-enabled JBAC system. Then, we propose an alternating algorithm to solve the formulated MOOP and obtain the sub-optimal solutions of the active beamforming vector $\mathbf{w}$ and the reflecting candidate set $\pmb{\Psi}$. 

\subsection{Problem Formulation}\label{PM}
In order to comprehensively strike the tradeoff between the performance of the primary communication and the backscatter communication, we formulate a MOOP that jointly maximizes the user average transmission rate of the primary communication and the goodput of the backscatter communication. The target of the formulated MOOP is to obtain the reflecting candidate set at the RIS and the active beamforming at the BS. The MOOP is capable of achieving the Pareto-optimal solution and enhancing the desired performance of the primary communication and the backscatter communication simultaneously. The above MOOP can be  formulated as

\begin{equation}
\begin{split}
 \left(\textbf{P1}\right): \;\;\;
\mathop {\max:}_{\pmb{\Psi}, \mathbf{w}} & R_{u}\left(\pmb{\Psi}, \mathbf{w}\right), \\
\mathop {\max : }_{\pmb{\Psi}, \mathbf{w}}& \bar{G}_{b}\left(\pmb{\Psi}, \mathbf{w}\right)\\
\rm{s.t. } &~\|\mathbf{w}\|^{2} \leq P_{\max }, \\ 
&~\left|\left(\pmb{\Phi}_{k}\right)_{n, n}\right|=1, \forall k , \forall n,
\end{split}
\end{equation}
where the first constraint indicates that the transmit power at the BS is smaller than the predefined threshold $P_{\max }$, the second one aims to satisfy that the RIS works in the passive mode and consumes no power. Clearly, the solution of $\left(\textbf{P1}\right)$ leads to the maximization of the joint performance of the proposed JBAC system. 


In the formulated MOOP, the Pareto-optimality is the most widely used solution, which relates to the so-called Pareto-frontier of the problem \cite{jing20}. One of the most popular methods to achieve the Pareto-optimality solution is the scalarization approach that transforms the MOOP into a single-objective optimization problem (SOOP) based on the preferential information related to the objectives. In this work, we obtain the Pareto optimal solution of $\left(\textbf{P1}\right)$ by adopting the weighted sum scalarization method. With $0 \leq \lambda_0 \leq 1$, the MOOP can be reformulated as
\begin{equation}
\begin{split}
 \left(\textbf{P2}\right): \;\;\;
\mathop {\max : }_{\pmb{\Psi}, \mathbf{w}} &  \lambda_0 \bar{{G}}_{b}\left(\pmb{\Psi}, \mathbf{w}\right)+(1-\lambda_0) R_{u}\left(\pmb{\Psi}, \mathbf{w}\right) \\
\rm{s.t. }: &~\|\mathbf{w}\|^{2} \leq P_{\max }, \\ 
&~\left|\left(\pmb{\Phi}_{k}\right)_{n, n}\right|=1, \forall k , \forall n.
\end{split}
\end{equation}
The weighting parameter $\lambda_0$ is a predefined value that describes the importance of $R_{u}$ and $G_b$. Due to the unit-modulus constraint on each reflecting element, the formulated SOOP is a non-convex optimization problem.
In the following, we present the solution of $(\textbf{P2})$.

\subsection{Alternative Solution} \label{AS}
The coupled relationship between $\mathbf{w}$ and $\pmb{\Psi}$ motivates us to take the alternating optimization method to find the stationary solution of $\left(\textbf{P2}\right)$. In the following, we first optimize $\mathbf{w}$ by fixing $\pmb{\Psi}$ and then update $\pmb{\Psi}$ by fixing $\mathbf{w}$. The sub-optimal solution of $ \left(\textbf{P2}\right)$ can be obtained by performing the two optimization processes iteratively until meeting the halting criterion. 

 \textit{1) Active Beamforming Design:}
With given $\pmb{\Psi}$,  $\left(\textbf{P2}\right)$ can be rewritten as
\begin{equation} 
\begin{split}
 \left(\textbf{P3}\right): \;\;\;
\mathop {\max : } \limits_{\mathbf{\mathbf{w}}} &~ \lambda_0 \bar{G}_{b}\left(\mathbf{w}\right)+(1-\lambda_0)R_{u}\left(\mathbf{w}\right)\\
\rm{s.t. }:  &~\|\mathbf{w}\|^{2} \leq P_{\max }. \\
\end{split}
\end{equation}
For illustration purposes, we define the feasible set of $\mathbf{w}$ as $\mathcal{W} = \{\mathbf{w} \in \mathbb{C}^{N_t \times 1}| \|\mathbf{w}\|^{2} \leq P_{\max }\}$. Then, we adopt the projected gradient method to solve $\left(\textbf{P3}\right)$. We first calculate the gradient of $f\left(\mathbf{w}\right) = \lambda_0 \bar{G}_{b}\left(\mathbf{w}\right)+(1-\lambda_0) R_{u}\left(\mathbf{w}\right)=\lambda_0 \bar{G}_{b}\left(\mathbf{w}\right)+\frac{1-\lambda_0}{M} \sum_{m=1}^MR_{m}\left(\mathbf{w}\right)$ with respect to $\mathbf{w}$ as 
\begin{equation}\label{NF}
\nabla_\mathbf{w}f\left(\mathbf{w}\right) = \lambda_0 \nabla_\mathbf{w}\bar{G}_{b}\left(\mathbf{w}\right)+\frac{1-\lambda_0}{M} \sum_{m=1}^M\nabla_\mathbf{w}R_{m}\left(\mathbf{w}\right), 
\end{equation}
which is the gradient sum of the goodput and the user average transmission rate in terms of $\mathbf{w}$. 

Specifically, $\nabla_\mathbf{w}\bar{G}_{b}\left(\mathbf{w}\right)$ can be calculated as
\begin{equation}\label{NG}
\begin{split}
&\nabla_\mathbf{w}\bar{G}_{b}\left(\mathbf{w}\right) = -\frac{\log _{2} K}{T_s} \mathbb{E}_{s(l)}[\nabla_\mathbf{w}\bar{P}_{e}\left(\mathbf{w}\right)]\\
& = -\frac{1}{T_s K} \mathbb{E}_{s(l)}[\sum_{k=1}^{K} \sum_{\hat{k}=1, \hat{k} \neq k}^{K} D_{\mathrm{HD}}\left({k}, {\hat{k}}\right) \nabla_\mathbf{w}P_{e}(k \rightarrow \hat{k},\mathbf{w})],
\end{split}
\end{equation}
where $\nabla_\mathbf{w}P_{e}(k \rightarrow \hat{k},\mathbf{w})$ is given as
\begin{equation}
\begin{split}
&\nabla_\mathbf{w}P_{e}(k \rightarrow \hat{k},\mathbf{w})= \nabla_\mathbf{w}\mathrm{Q}\left(\sqrt{\frac{D_{\mathrm{ED}}(k \rightarrow \hat{k}, \mathbf{w})^{2}}{2N_0}}\right)\\
& = -{\frac{1}{2}}\sqrt{\frac{1}{\pi N_0 D_{\mathrm{ED}}(k \rightarrow \hat{k}, \mathbf{w})^2}} \\
&\times \exp \left(-\frac{ D_{\mathrm{ED}}(k \rightarrow \hat{k}, \mathbf{w})^2}{4 N_0}\right) \tilde{\mathbf{H}}^H_{k,\hat{k}}\tilde{\mathbf{H}}_{k,\hat{k}}\mathbf{w},
\end{split}
\end{equation}
with $\tilde{\mathbf{H}}_{k,\hat{k}} = \mathbf{H}_{b} \pmb{\Phi}_{k} \mathbf{H}_{1}s(l)- \mathbf{H}_{b} \pmb{\Phi}_{\hat{k}} \mathbf{H}_{1}s(l)$. 

The gradient of the achievable rate at the $m$-th user with respect to $\mathbf{w}$ is 

\begin{equation}\label{NR}
\nabla_\mathbf{w}R_{m}\left(\mathbf{w}\right) = \frac{1}{T_{s}K} \sum_{k=1}^K\frac{2\tilde{\mathbf{h}}_m^H(\pmb{\Phi}_k)\tilde{\mathbf{h}}_m(\pmb{\Phi}_k)\mathbf{w}}{\ln 2 \sigma_{m}^2(1+\gamma_{m}(\pmb{\Phi}_k))},
\end{equation}
with $\tilde{\mathbf{h}}_m(\pmb{\Phi}_k) = \mathbf{h}_{2,m}^H \pmb{\Phi}_k \mathbf{H}_{1}+\mathbf{h}_{d,m}^H$. 
Substituting \eqref{NG} and \eqref{NR} into \eqref{NF}, we can get the gradient of $f\left(\mathbf{w}\right)$ with respect to $\mathbf{w}$. Then, a stationary solution of $ \left(\textbf{P3}\right)$ can be obtained by the projected gradient method. Let $\mathbf{w}^{(t)}$ denote the active beamformer in the $t$-th iteration and $\alpha_{\mathbf{w}}$ be the step size corresponding to $\mathbf{w}$. Given the privious iteration value $\mathbf{w}^{(t-1)}$, the objective function can be increased by moving along its gradient direction $\nabla_\mathbf{w}f\left(\mathbf{w}\right)$
 with step size $\alpha_{\mathbf{w}}$. We can get $ \hat{\mathbf{w}}^{(t)}$ as
 \begin{equation}\label{hatw}
 \hat{\mathbf{w}}^{(t)} \triangleq \mathbf{w}^{(t-1)}+\alpha_{\mathbf{w}} \nabla_{\mathbf{w}} f\left(\mathbf{w}^{(t-1)}\right). 
 \end{equation}
 Then, we project $\hat{\mathbf{w}}^{(t)}$ onto $\mathcal{W}$ to update $\mathbf{w}^{(t)}$ given by
 \begin{equation}\label{PW}
 \mathbf{w}^{(t)}=\sqrt{P_{\max }} \hat{\mathbf{w}}^{(t)} / \max \left\{\left\|\hat{\mathbf{w}}^{(t)}\right\|, \sqrt{P_{\max}}\right\} .	
 \end{equation}

The steps of the active beamforming design are presented in $\textbf{Algorithm 1}$.

\begin{algorithm}[t]
\caption{Projected Gradient Method for Optimizing $\mathbf{w}$}
\label{10}
\begin{algorithmic}[1]
\STATE {Set $t=1$, Initialize $\mathbf{w}^{(0)}$ to feasible values.}
\STATE{Get the ascent gradient according to \eqref{NF}.}
\STATE{Update $\hat{\mathbf{w}}^{(t)}$ according to \eqref{hatw}.}
\STATE{Project $\hat{\mathbf{w}}^{(t)}$ onto its feasible set $\mathcal{W}$ according to \eqref{PW} and let $t+1 \leftarrow t$.}
\STATE{Repeat steps 2-4 until the halting criterion is met.}
\STATE{Output $\mathbf{w}^{(t)}$.}
\end{algorithmic}
\end{algorithm}

\textit{2) Reflecting Pattern Design:}
Based on the optimized  $\mathbf{w}$, the optimization problem $\left(\textbf{P2}\right)$ can be rewritten as
\begin{equation}
\begin{split}
 \left(\textbf{P4}\right): \;\;\;
\mathop {\max : }_{\pmb{\Psi}} & \lambda_0 \bar{G}_{b}\left(\pmb{\Psi}\right)+(1-\lambda_0) R_{u}\left(\pmb{\Psi}\right) \\
\rm{s.t. }:&~\left|\left(\pmb{\Phi}_{k}\right)_{n, n}\right|=1, \forall k , \forall n.
\end{split}
\end{equation}

For convenience, we further express the reflecting candidate set as a diagonal matrix that includes all the reflecting patterns as
\begin{equation}
\pmb{\Xi}=
\begin{bmatrix}
    ~\pmb{\Phi}_1 & \textbf{0} &\cdots&\textbf{0}\\
    ~\textbf{0} & ~\pmb{\Phi}_2& \cdots& \textbf{0}~ \\
~\vdots& \vdots &  \ddots& \vdots~\\
    ~\textbf{0} &\textbf{0} &\cdots &\pmb{\Phi}_{K}\\
\end{bmatrix}\in \mathbb{C}^{KN\times KN}.
\end{equation}
We set $\pmb{\uppsi} = \operatorname{diag}(\pmb{\Xi})$ and transform the non-convex constraint $\left|\left(\pmb{\Phi}_{k}\right)_{n, n}\right| = \left|\pmb{\uppsi}_{N(k-1)+n}\right|=1$ to an equivalent constraint as
\begin{equation}
\begin{aligned}
&\left\{\pmb{\uppsi} \in \mathbb{C}^{N K \times 1}:\left|\pmb{\uppsi}_{i}\right|, \forall i=1, \ldots, N K\right\} \\
&=\left\{\pmb{\uppsi}\in \mathbb{C}^{N K \times 1}: \operatorname{tr}\left(\pmb{\uppsi}\pmb{\uppsi}^H\right)=N K,\|\pmb{\uppsi}\|_{\infty} \leq 1\right\}. 
\end{aligned}
\end{equation}
However, $l_{\infty}$ is non-differentiable. One of the feasible solutions is to replace $l_{\infty}$ by the $l_p$ approximation with a gradually increased large $p$ during the optimization process. Therefore, the optimization problem $ \left(\textbf{P4}\right)$ can be re-expressed as

 \begin{equation}
\begin{split}
 \left(\textbf{P5}\right): \;\;\;
\mathop {\max : }_{\pmb{\uppsi}} & \lambda_0 \bar{G}_{b}\left(\pmb{\uppsi}\right)+(1-\lambda_0) R_{u}\left(\pmb{\uppsi}\right) \\
\rm{s.t. }:&~\operatorname{tr}\left(\pmb{\uppsi}\pmb{\uppsi}^H\right)=N K,\\
&~\|\pmb{\uppsi}\|_{p} \leq 1, p \rightarrow \infty.
\end{split}
\end{equation}

The optimization problem $\left(\textbf{P5}\right)$ can be solved by the projected gradient method in \cite{Guo2020} once the gradient of $g\left(\pmb{\uppsi}\right) = \lambda_0 \bar{G}_{b}\left(\pmb{\uppsi}\right)+(1-\lambda_0) R_{u}\left(\pmb{\uppsi}\right)=\lambda_0\bar{G}_{b}\left(\pmb{\uppsi}\right) +\frac{1-\lambda_0}{M} \sum_{m=1}^MR_{m}\left(\pmb{\uppsi}\right)$ with respect to $\pmb{\uppsi}$ can be derived in its closed-form. It is obvious that $\nabla_{\pmb{\uppsi}} g\left(\pmb{\uppsi}\right)$ is decided by $\nabla_{\pmb{\uppsi}} \bar{G}_{b}\left(\pmb{\uppsi}\right)$ and $\nabla_{\pmb{\uppsi}} R_{m}\left(\pmb{\uppsi}\right)$.   $\nabla_{\pmb{\uppsi}} \bar{G}_{b}\left(\pmb{\uppsi}\right)$ is calculated as

\begin{equation}\label{NP}
\begin{split}
\nabla_{\pmb{\uppsi}} \bar{G}_{b}\left(\pmb{\uppsi}\right) &= \left[\nabla_{\pmb{\upphi}_1} \bar{G}_{b}\left(\pmb{\upphi}_1\right)^T, \right.\\
&\left.\nabla_{\pmb{\upphi}_2} \bar{G}_{b}\left(\pmb{\upphi}_2\right)^T, \cdots, \nabla_{\pmb{\upphi}_K} \bar{G}_{b}\left({\pmb{\upphi}_K}\right)^T\right]^T,
\end{split}
\end{equation}
with

\begin{equation}\label{NPk}
\begin{split}
&\nabla_{\pmb{\upphi}_k} \bar{G}_{b}\left({\pmb{\upphi}_k}\right) = -\frac{\log _{2} K}{T_s}\mathbb{E}_{s(l)}[\nabla_{\pmb{\upphi}_k} \bar{P}_{e}]\\
& = -\frac{1}{T_s K} \mathbb{E}_{s(l)}[\sum_{\hat{k}=1, \hat{k} \neq k}^{K} D_{\mathrm{HD}}\left({k}, {\hat{k}}\right) \nabla_{\pmb{\upphi}_k}P_{\mathrm{e}}(k \rightarrow \hat{k})]\\
&-\frac{1}{T_s K} \mathbb{E}_{s(l)}[\sum_{\hat{k}=1, \hat{k} \neq k}^{ K} D_{\mathrm{HD}}\left(\hat{k}, {{k}}\right) \nabla_{\pmb{\upphi}_k} P_{\mathrm{e}}(\hat{k} \rightarrow k)]. 
\end{split}
\end{equation}

By utilizing the third equation of \eqref{DED}, $\nabla_{\pmb{\upphi}_k}P_{\mathrm{e}}(k \rightarrow \hat{k})$ and $\nabla_{\pmb{\upphi}_k} P_{\mathrm{e}}(\hat{k} \rightarrow k)$ can be easily obtained as

\begin{equation}\label{Nk}
\begin{split}
&\nabla_{\pmb{\upphi}_k}P_{\mathrm{e}}(k \rightarrow \hat{k}) = -{\frac{1}{2}}\sqrt{\frac{1}{\pi N_0 D_{\mathrm{ED}}(k \rightarrow \hat{k})^2}} \\
&\times \exp \left(-\frac{ D_{\mathrm{ED}}(k \rightarrow \hat{k})^2}{4 N_0}\right) \bar{\mathbf{H}}^H\bar{\mathbf{H}}\left(\pmb{\upphi}_k-\pmb{\upphi}_{\hat{k}}	\right),
\end{split}
\end{equation}

and 

\begin{equation}\label{Nkh}
\begin{split}
&\nabla_{\pmb{\upphi}_k}P_{\mathrm{e}}(\hat{k} \rightarrow k) = {\frac{1}{2}}\sqrt{\frac{1}{\pi N_0 D_{\mathrm{ED}}(\hat{k} \rightarrow k)^2}} \\
&\times \exp \left(-\frac{ D_{\mathrm{ED}}(\hat{k} \rightarrow k)^2}{4 N_0}\right) \bar{\mathbf{H}}^H\bar{\mathbf{H}}\left(\pmb{\upphi}_{\hat{k}}-\pmb{\upphi}_k\right).
\end{split}
\end{equation}

By substituting \eqref{Nk} and \eqref{Nkh} into \eqref{NPk}, we can get $\nabla_{\pmb{\upphi}_k} \bar{G}_{b}\left({\pmb{\upphi}_k}\right)$ and $\nabla_{\pmb{\uppsi}} \bar{G}_{b}\left(\pmb{\uppsi}\right)$. Similarly, the gradient of $R_{m}\left(\pmb{\uppsi}\right)$,  $\nabla_{\pmb{\uppsi}} R_{m}\left(\pmb{\uppsi}\right)$, can be obtained as

\begin{equation}\label{NPR}
\begin{split}
\nabla_{\pmb{\uppsi}} R_{m}\left(\pmb{\uppsi}\right) &= \left[\nabla_{\pmb{\upphi}_1} R_{m}\left(\pmb{\upphi}_1\right)^T, \right.\\
&\left.\nabla_{\pmb{\upphi}_2} R_{m}\left(\pmb{\upphi}_2\right)^T, \cdots, \nabla_{\pmb{\upphi}_K} R_{m}\left({\pmb{\upphi}_K}\right)^T\right]^T,
\end{split}
\end{equation}

 with

\begin{equation}\label{NRm}
\nabla_{\pmb{\upphi}_k} R_{m}\left(\pmb{\upphi}_k\right) = \frac{1}{T_{s}K} \frac{\nabla_{\pmb{\upphi}_k}\gamma_{m}(\pmb{\upphi}_k)}{\ln 2 \left(1+\gamma_{m}(\pmb{\upphi}_k)\right)},
\end{equation}

and 
 $\nabla_{\pmb{\upphi}_k}\gamma_{m}(\pmb{\upphi}_k)$ as
\begin{equation}\label{NGk}
\begin{split}
\nabla_{\pmb{\upphi}_k}\gamma_{m}(\pmb{\upphi}_k) &= \frac{\nabla_{\pmb{\upphi}_k}|(\mathbf{h}_{2,m}^H \pmb{\Phi}_k \mathbf{H}_{1}+\mathbf{h}_{d,m}^H)\mathbf{w}|^{2}}{\sigma_{m}^2}\\
& = \frac{\nabla_{\pmb{\upphi}_k}|\mathbf{h}_{2,m}^H \operatorname{diag}\left(\mathbf{H}_{1}\mathbf{w}\right)\pmb{\upphi}_k+\mathbf{h}_{d,m}^H\mathbf{w}|^{2}}{\sigma_{m}^2}\\
& =  \frac{\nabla_{\pmb{\upphi}_k}|\bar{\mathbf{h}}_m\pmb{\upphi}_k +c_{d,m}|^{2}}{\sigma_{m}^2}\\
& = \frac{2\bar{\mathbf{h}}_m^H\left(\bar{\mathbf{h}}_m\pmb{\upphi}_k +c_{d,m}\right)}{\sigma_{m}^2},
\end{split}
\end{equation}
where we have $\bar{\mathbf{h}}_m = \mathbf{h}_{2,m}^H {\operatorname{diag}}\{\mathbf{H}_{1}\mathbf{w}\}$ and $c_{d,m} = \mathbf{h}_{d,m}^H\mathbf{w}$. With the obtained $\nabla_{\pmb{\uppsi}} \bar{G}_{b}\left(\pmb{\uppsi}\right)$ and $\nabla_{\pmb{\uppsi}} R_{m}\left(\pmb{\uppsi}\right)$, the gradient of the objective function $\nabla_{\pmb{\uppsi}} g\left(\pmb{\uppsi}\right)$ in $ \left(\textbf{P5}\right)$ can be obtained. 

Based on the available gradient of the objective function, the solution of $\left(\textbf{P5}\right) $ can be founded iteratively as in \cite{Guo2020}. Firstly, we adopt the barrier method and cope with the non-negative constraint $\|\pmb{\uppsi}\|_{p} \leq 1$, $p \rightarrow \infty$ by the logarithmic barrier function $V(\cdot)$. 

\begin{align}
V\left( u \right) = \left\{ {\begin{array}{*{20}{c}}
{ \frac{1}{q}\ln \left( u \right),}&{u > 0}\\
{-\infty ,}&{u \le 0,}
\end{array}} \right.
\end{align}
where $q$ is the penalty parameter. Secondly, we derive the search direction according to \eqref{Grad}, and then project it onto the tangent plane of $\operatorname{tr}\left(\pmb{\uppsi}\pmb{\uppsi}^H\right)=N K$. Finally, we update $\pmb{\uppsi}$ through the projected gradient method. Based on the proposed design of the active beamforming and the reflecting patterns, the system can be alternatively optimized.

\begin{equation}\label{Grad}
	\begin{split}
	L(\pmb{\uppsi})&=
	\nabla_{\pmb{\uppsi}} g\left(\pmb{\uppsi}\right) + \nabla_{\pmb{\uppsi}} V(1-||\pmb{\uppsi}||_p)\\&=\nabla_{\pmb{\uppsi}}g\left(\pmb{\uppsi}\right)-\frac{\|\pmb{\uppsi}\|_p^{1-p} \mathbf{p}_{\pmb{\uppsi}}}{q\left(1-\|\pmb{\uppsi}\|_p\right)},
	\end{split}
\end{equation}
with $\textbf{p}_{\pmb{\uppsi}}={\left[ {{{\uppsi} _1} \cdot {{\left| {{{\uppsi} _1}} \right|}^{p - 2}},{{\uppsi} _2} \cdot {{\left| {{{\uppsi} _2}} \right|}^{p - 2}},...,{{\uppsi} _{{K}N}} \cdot {{\left| {{{\uppsi}_{{K}N}}} \right|}^{p - 2}}} \right]^T}$.

\begin{remark}
It is worth noting that the reflecting patterns will be assigned to different indices and carry additional information. According to \eqref{pe}, the BER performance is related to the Hamming distances $D_{\mathrm{HD}}\left({k}, {\hat{k}}\right)$ and the Euclidean distances $D_{\mathrm{ED}}(k \rightarrow \hat{k})$. The mapping rule directly affects the Hamming distances. Therefore, the BER of the backscatter communication system can be further reduced by adopting a suitable mapping scheme between the reflecting patterns and the indices. One of the typical methods is Pseudo-Gray mapping \cite{zeger1990pseudo}, which can be easily implemented through a binary switching algorithm (BSA) \cite{Guo2016}. Specifically, the BSA starts with an initial mapping, and then switches the bit labels of any two reflecting patterns to generate a new mapping. By comparing the original one and the switched one in terms of BER, the mapping with better BER performance will be chosen. Thus, the Pseudo-Gray mapping can be obtained by repeating the processes above until all the index labels are exchanged. In order to validate the effectiveness of the BSA on the proposed system, we also compare the performance of the system with BSA and the one without BSA in the following.  
\end{remark}

\subsection{Complexity Analysis}

In this subsection, we analyze the computational complexity of the proposed alternating algorithm, which mainly originates from the design of the active beamforming and the reflecting patterns. In addition, the computational complexity of BSA is also presented.

\subsubsection{Complexity of the Active Beamforming Design} The computational complexity is induced by the gradient calculation. The computational complexity of $\tilde{\mathbf{H}}^H_{k,\hat{k}}\tilde{\mathbf{H}}_{k,\hat{k}}\mathbf{w}$ is ${\mathcal{O}}[K^2{N}_r(N^2+N{N}_t+{{N}_t}^2)]$ and $\tilde{\mathbf{h}}_m^H(\pmb{\Phi}_k)\tilde{\mathbf{h}}_m(\pmb{\Phi}_k)\mathbf{w}$ is ${\mathcal{O}}[KM(N^2+N{N}_t+{{N}_t}^2)]$, respectively. Thus, the computational complexity of the active beamforming design is 
\begin{equation}
\begin{split}
&{\mathcal{C}}_w={\mathcal{O}}[KI_w(K{N}_rN^2 +K{N}_rN{N}_t +K{N}_r{{N}_t}^2 \\
&+MN^2+MNN_{t}+M{{N}_t}^2)], 
\end{split}
\end{equation}
where $I_{w}$ denotes the number of iterations in the active beamforming design.

\subsubsection{Complexity of the Reflecting Pattern Design} The computational complexity is dominated by the gradient in \eqref{NP} and \eqref{NPR}. $\bar{\mathbf{H}}^H\bar{\mathbf{H}}(\pmb{\upphi}_k-\pmb{\upphi}_{\hat{k}})$ and  $\bar{\mathbf{H}}^H\bar{\mathbf{H}}(\pmb{\upphi}_{\hat{k}}-\pmb{\upphi}_k)$ in \eqref{NP} introduce the computational complexity of ${\mathcal{O}}[K^2(N{N}_t+N^2{{N}_r})]$, and the computational complexity of $\bar{\mathbf{h}}_m^H\bar{\mathbf{h}}_m\pmb{\upphi}_k$ in \eqref{NPR} is ${\mathcal{O}}[KMN(N+{N}_t)]$. Thus, the computational complexity of the reflecting pattern design is 
\begin{equation}
\begin{split}
&{\mathcal{C}}_{\phi}={\mathcal{O}}[KI_{\phi} (KN{N}_t+KN^2{N}_r+MN^2+MNN_{t})], 
\end{split}
\end{equation}
where $I_{\phi}$ denotes the number of iterations in the reflecting pattern design.
\subsubsection{Complexity of BSA}The complexity of BSA is mainly dominated by $P_e$, and it can be computed as 
\begin{equation}
\begin{split}
&{\mathcal{C}}_{BSA}={\mathcal{O}}[K^4 (N^2 N_r + N N_t)]. 
\end{split}
\end{equation}

Let $I$ stand for the required number of iterations in the alternating algorithm, the overall complexity of the proposed algorithm can be calculated as 
\begin{equation}
\begin{split}
&{\mathcal{C}}=I({\mathcal{C}}_{w}+{\mathcal{C}}_{\phi})+{\mathcal{C}}_{BSA}. 
\end{split}
\end{equation}

\section{numerical results}\label{SR}
In this section, the performance of the RIS-enabled JBAC system and the effectiveness of the developed alternating algorithm are evaluated. For simplicity, we term the objective function $\lambda_0 \bar{{G}}_{b}\left(\pmb{\Psi}, \mathbf{w}\right)+(1-\lambda_0) R_{u}\left(\pmb{\Psi}, \mathbf{w}\right)$ as ``weighted system sum rate". The algorithm to design the active beamforming vector $\mathbf{w}$ is called ``Active Beamforming" and the algorithm to make the reflecting candidate set $\pmb{\Psi}$ is called ``Passive Beamforming". We first investigate the convergence behaviors of the proposed alternating algorithm. The comparison between the proposed algorithm and the exhaustive search (ES) algorithm is also presented. Then, we investigate the effects of $\lambda_0$, $K$, and $N$ on the weighted system sum rate performance. We also compare the proposed algorithm with the existing algorithms. Furthermore, the impacts of imperfect CSI and discrete phase shifts are also presented. We set the noise power for each terminal to be the same for simplify, i.e., $N_0=\sigma_{m}^2$ $(m=1, \cdots,M)$ \cite{Luo2023,Liu2021}. In the following, all numerical results are investigated under standard Rayleigh fading channels. Other required parameters are presented in Table I. 
\begin{table}[ht]   
\begin{center}   
\caption{Simulation Setups}  
\label{table:1} 
\begin{tabular} {|c|c|}
\hline   \textbf{Parameter} & \textbf{Values}\\   
\hline  Transceiver antenna configuration at the BS ($N_t$, $N_r$) & ($2$, $4$) \\
\hline   Number of reflecting elements $N$ & $(2, 10, 20)$ \\  
\hline   Number of reflecting patterns $K$ & $(4, 8)$\\      
\hline  Number of users $M$ & $4$\\ 
\hline   Symbol duration $T_s$ & $50~\mu s$\\
\hline   The maximiun transmit power at the BS $P_{\max}$ & $0$ dB\\
\hline
\end{tabular}    
\end{center}   
\end{table}

\begin{figure*}
	\setlength{\abovecaptionskip}{-5pt}
	\setlength{\belowcaptionskip}{-10pt}
	\centering
	\begin{minipage}[t]{0.33\linewidth}
		\centering
		\includegraphics[width=2.56in]{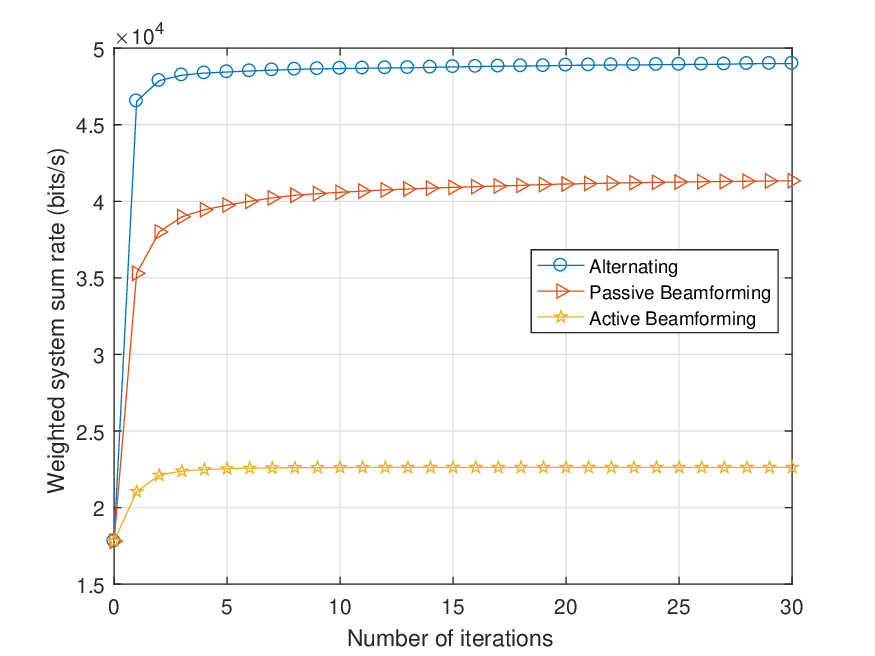}
		\label{figure1}
	\end{minipage}%
	\begin{minipage}[t]{0.33\linewidth}
		\centering
		\includegraphics[width=2.6in]{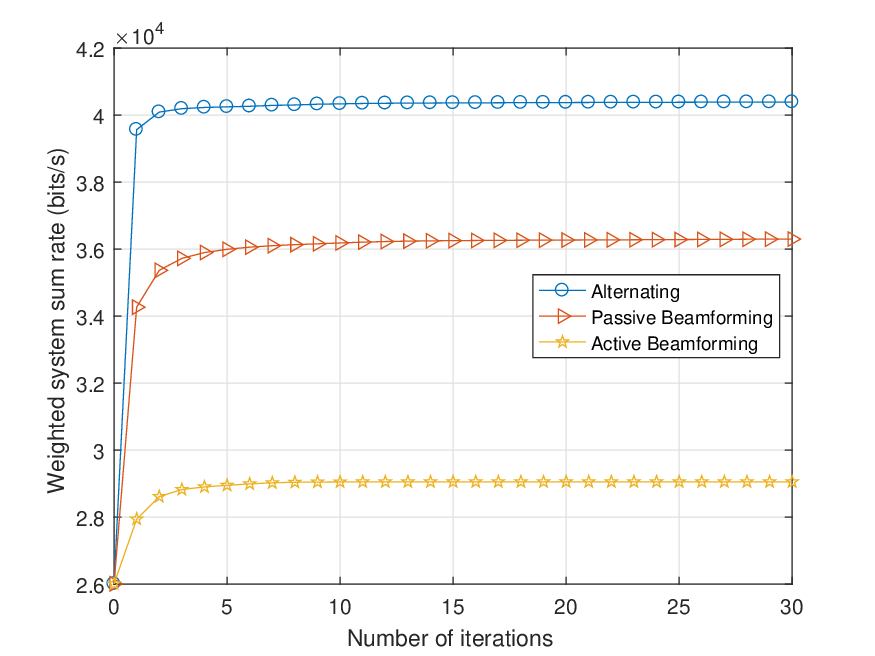}
		\label{figure2}
	\end{minipage}
	\begin{minipage}[t]{0.33\linewidth}
		\centering
		\includegraphics[width=2.6in]{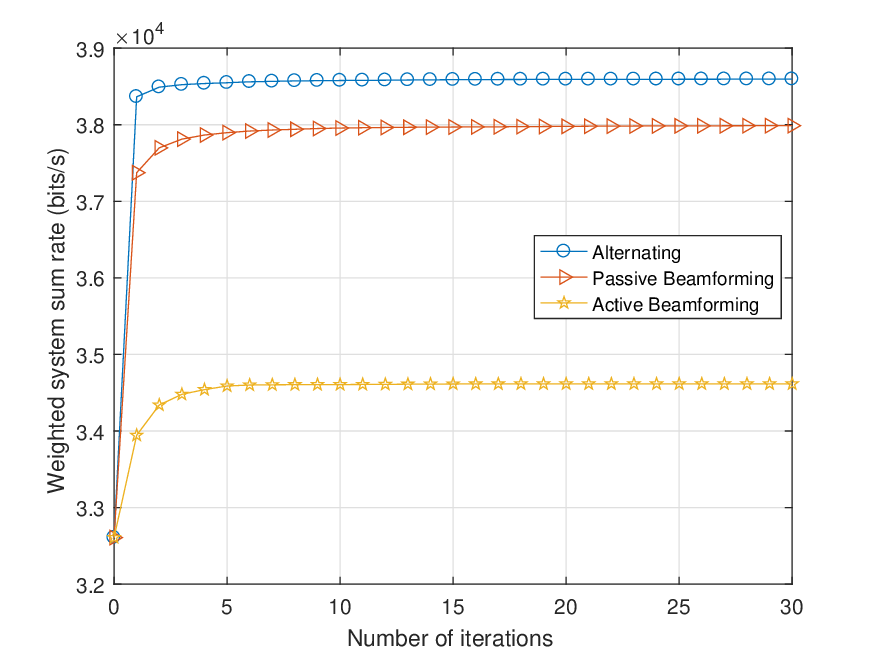}
		\label{figure3}
	\end{minipage}
 \caption{Convergence of the proposed iterative beamforming method in (a) $\lambda_0=0$, (b) $\lambda_0=0.5$, (c) $\lambda_0=1$ at $\mathrm{SNR}=-10$ dB.}
 \label{Convergence}
\end{figure*}

Firstly, we study the effectiveness of the proposed alternating algorithm in the considered JBAC system. The convergence behaviors of all the proposed algorithms and the performance in terms of weighted system sum rate are given in Fig. \ref{Convergence}. It is shown that, all algorithms converge very fast for different $\lambda_0$. It means that the proposed algorithm offers an efficient way to optimize the active beamforming and the passive beamforming patterns. The alternating algorithm achieves the best performance among the three algorithms, followed by the passive beamforming and the active beamforming. The small gap of the goodput achieved by the alternating algorithm and the passive beamforming in Fig. \ref{Convergence}(c) indicates that we can solely optimize the reflecting patterns to save some computational resources when we only only care about the performance of the backscatter communication. However, when the goodput and the user average transmission rate have the same weight or we aim to improve the user average transmission rate, optimizing the active beamforming and the reflecting patterns iteratively is critical. On the other hand, the performance of the achieved weighted system sum rate at $\lambda_0 = 0$ is better than that at $\lambda_0 = 0.5$ and $\lambda_0 = 1$. This is because the user average transmission rate of the primary communication is greater than the goodput of the backscatter communication at $\mathrm{SNR}=-10$ dB.

\begin{figure}[t]
  \centering
  \includegraphics[width=0.48\textwidth]{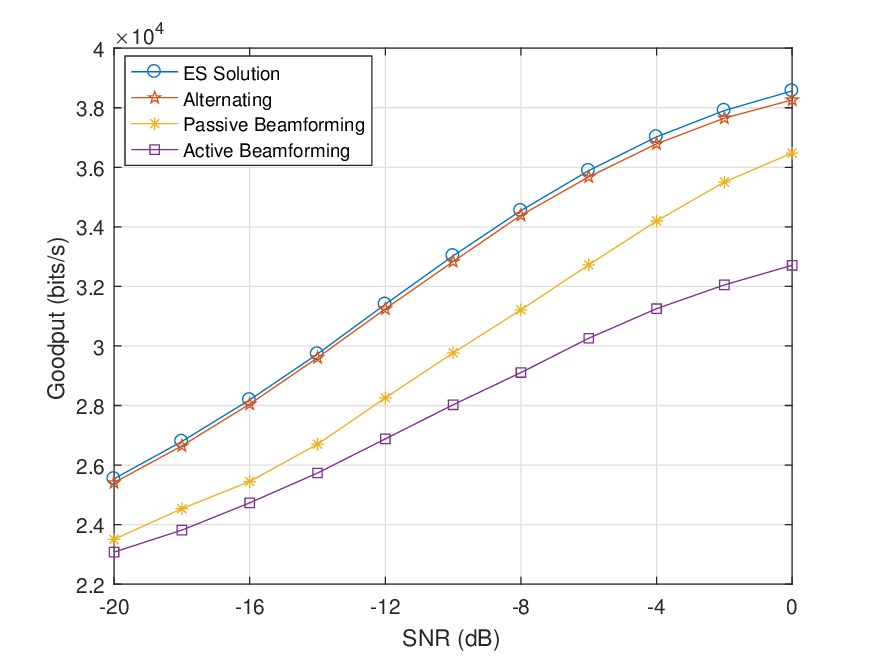}
  \caption{Goodput comparison between the proposed algorithms and ES algorithm versus $\mathrm{SNR}$ with $N=2$ and $b$=2.} 
  \label{ES}
\end{figure}

To investigate the effectiveness of the proposed algorithm, we compare it with ES algorithm in terms of goodput performance in Fig. \ref{ES}. We set $C_{\mathrm{ESR}}$ and $C_{\mathrm{ESW}}$ as the feasible set of phase shifts and active beamforming, respectively. Thus, the ES algorithm needs to compute $\binom{C_\mathrm{ESR}^N}{K} C_{\mathrm{ESW}}^{N_t}$ times of calculations of $\bar{P}_{e}$, whose complexity increases exponentially with the number of $N_t$ and $N$. Furthermore, since searching continuous reflecting patterns and active beamforming requires huge time and computational resources to achieve high accuracy, we adopt a discrete set of phase shifts as a substitution, i.e., $\left(\pmb{\Phi}_{k}\right)_{n, n} \in \left\{e^{\frac{j 2 \pi m}{2^b} }\right\}_{m=0}^{2^{b}-1}$, $\forall n \in\{1,2 \ldots, N\}$, where $b$ is the number of quantization bits. We set $N=2$ and $b=2$ for further reducing the complexity of ES algorithm in implementation. We can see that the performance of the proposed sub-optimal solution is very close to that of the ES algorithm. Thus, the proposed algorithm is more practical.

\begin{figure}[t]
  \centering
  \includegraphics[width=0.48\textwidth]{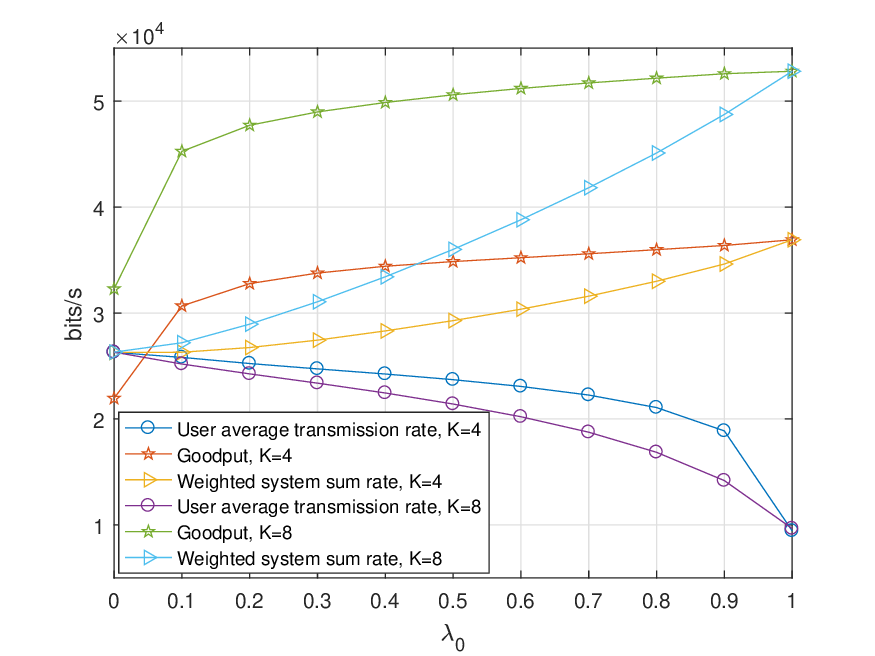}
  \caption{The user average transmission rate, goodput, and weighted system sum rate versus $\lambda_0$.} 
  \label{Weights}
\end{figure}

In Fig. \ref{Weights}, we further investigate the impact of $\lambda_0$ on the user average transmission rate of the primary communication system, the goodput of the backscatter communication system, and the weighted system sum rate of the entire system. As shown in Fig. \ref{Weights}, with the increase of $\lambda_0$, i.e., the goodput takes more weight, the performance of weighted system sum rate increases obviously. This is due to the fact that the user average transmission rate always being smaller than the goodput at $\mathrm{SNR}=-15$ dB. Moreover, when the weight of the goodput goes to be 1, there are obvious performance loss of the user average transmission rate. The increase of $\lambda_0$ can greatly improve the goodput of the backscatter communication system with $\lambda_0<0.1$. According to \eqref{GP}, the goodput is directly related to the BER of the backscatter system. When $\lambda_0$ increases from 0 to 0.1, the BER performance of the backscatter communication system is reduced obviously, so the goodput can be significantly improved. The BER performance has a lower bound, which leads to an upper bound of the goodput. Since the decrease of the BER performance becomes very limited when $\lambda_0>0.1$, the improvement of the goodput becomes marginal. These results also inspire us to choose a suitable $\lambda_0$ for different system objectives in practice. In addition, it can be found that the goodput and the weighted system sum rate at $K=8$ are greater than that at $K=4$. Although a higher value of $K$ may cause worse BER performance, it can also improve the upper bound of goodput in the backscatter communication for the proposed system. We can also observe that the increased reflecting candidate set size slightly impairs the primary communication system. Since the reflecting patterns that can achieve higher transmission rate have already been included at $K=4$, we can only add the reflecting patterns that achieve worse transmission rate performance when $K$ increases to 8. Thus, the transmission rate of the primary communication will be slightly degraded. 

\begin{figure}[t]
  \centering
  \includegraphics[width=0.48\textwidth]{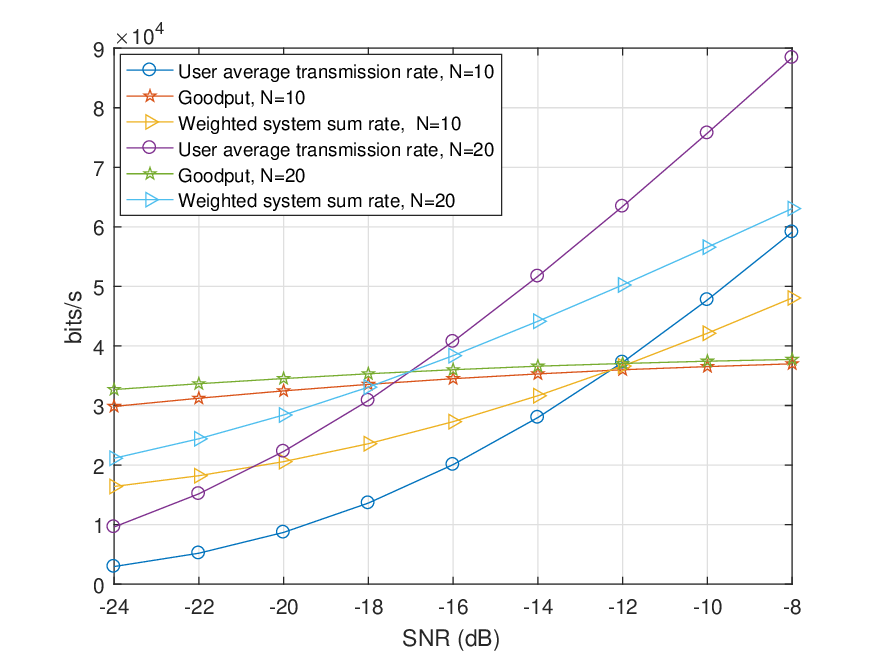}
  \caption{The user average transmission rate, goodput and weighted system sum rate versus $\mathrm{SNR}$ with $\lambda_0=0.5$.} 
  \label{DifferentN}
\end{figure}

In order to explore the effect of $N$ on system performance, we observe the performance of the backscatter communication system, the primary communication system, and the entire system with $N=10$ and $N=20$ in Fig. \ref{DifferentN}. It is seen that the user average transmission rate of the primary communication system can be significantly improved with the increase of $\mathrm{SNR}$ or $N$. When the user average transmission rate achieves $4\times10^{4}$ bits/s, around $4.8$ dB gains are obtained by increasing $N$ from 10 to 20. With the increase of $\mathrm{SNR}$, the goodput approaches its upper bound of the backscatter communication. The user average transmission rate exceeds the goodput when $\mathrm{SNR}=-12.3$ dB and $-17$ dB under $N=10$ and $N=20$, respectively. However, there is no obvious change in the goodput of the backscatter communication system with the increase of $\mathrm{SNR}$ and $N$. By combining the phenomena observed in Fig. \ref{DifferentN} and  Fig. \ref{Weights}, we can find that the performance of the RIS-enabled backscatter communication system is influenced by the size of the reflecting candidate set effectively. The above results give us some implementation remarks when we implement the RIS-enabled JBAC system in practice. The goodput of the backscatter communication can be improved by increasing the number of reflecting patterns, while the user average transmission rate of the primary communication can be enhanced by improving the number of the reflecting elements.

\begin{figure}[t]
  \centering
  \includegraphics[width=0.48\textwidth]{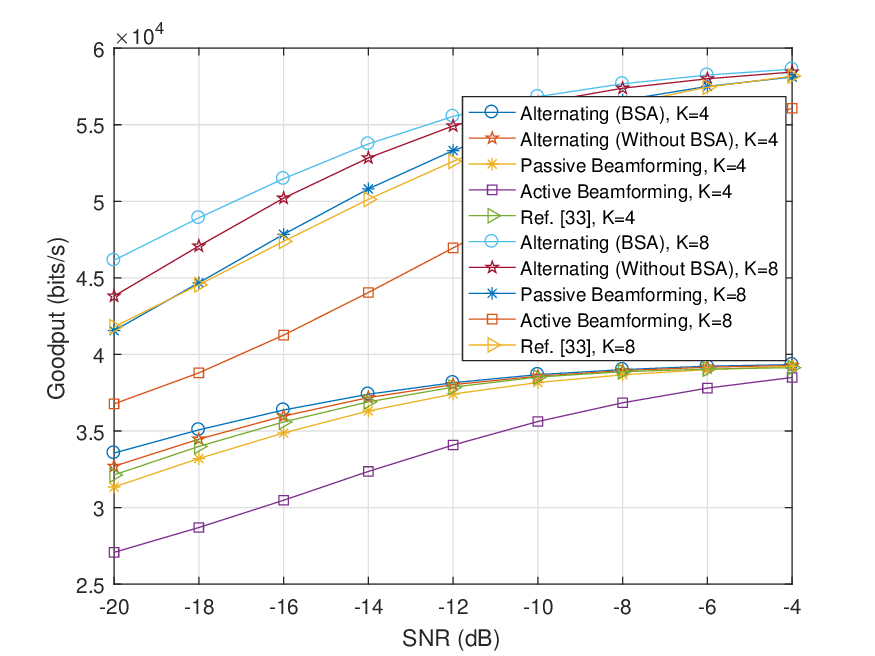}
  \caption{Goodput of RIS-enabled backscatter communication system versus $\mathrm{SNR}$ under different optimization schemes.} 
  \label{Backscatter}
\end{figure}

In the following, we evaluate the goodput performance of the RIS-enabled backscatter communication system under different optimization schemes. In Fig. \ref{Backscatter}, we set $\lambda_0=1$ and ignore the primary communication system. 
The alternating algorithm without the BSA, the active beamforming design, the passive beamforming design, and the algorithm with the scheme proposed in \cite{zhang2021reconfigurable} are severed as benchmarks. In\cite{zhang2021reconfigurable}, RIS is designed to carry additional information by multiplying the phase-shift matrix with the signals modulated by $M$-PSK modulation scheme. We set $M=K$ for comparison. The simulation results show that the proposed alternating algorithm with the BSA outperforms other benchmarks. The performance of the alternating algorithm without the BSA is little worse than that of the alternating algorithm with the BSA, and outperforms the performance of the scheme in \cite{zhang2021reconfigurable}. Performing the active beamforming or the passive beamforming independently sometimes achieves worse performance compared with the scheme in \cite{zhang2021reconfigurable}. In detail, when the goodput achieves $5\times10^{4}$ bits/s, the alternating algorithm with $K = 8$ can obtain about $3.7$ dB gains compared with the scheme in \cite{zhang2021reconfigurable}. 
Due to the fact that the multiplication operation on the phase shifts impairs the achieved performance gain provided by the RIS and the predefined symbol format is irrelevant to the CSI, the algorithm with the scheme in \cite{zhang2021reconfigurable} achieves worse performance compared with the proposed alternating algorithm.
 Moreover, the simulation results verify the importance of the reflecting pattern design, which is the same as the conclusion in Fig. \ref{Convergence}. 
 
\begin{figure}[t]
  \centering
  \includegraphics[width=0.48\textwidth]{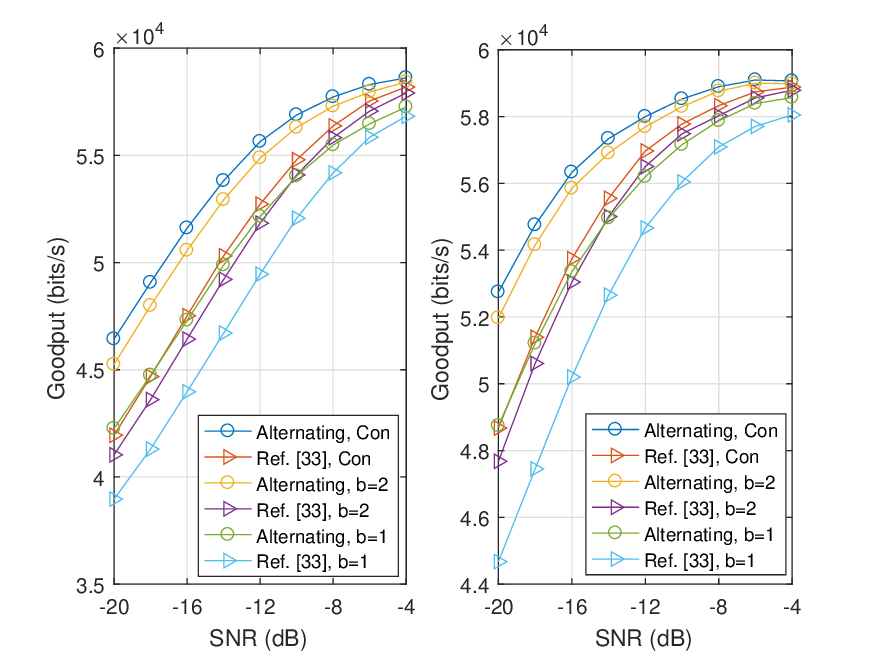}
  \caption{Goodput of the alternating algorithm and the scheme in [33] with $K=8$ for (a) $N=10$, (b) $N=20$.} 
  \label{Disphase}
\end{figure}

In the following, the effect of the discrete phase shifts on the system goodput performance is also examined with respect to $\mathrm{SNR}$ in Fig. \ref{Disphase}. As the number of quantization bits increases, the goodput of the proposed alternating algorithm and the scheme in \cite{zhang2021reconfigurable} can be obtained as in the continuous case. The proposed alternating algorithm and the scheme in \cite{zhang2021reconfigurable} with $b=2$ obtain almost the same performance compared to the continuous phase shifts. It inspires us to choose an appropriate $b$ to achieve the performance of the continuous phase shifts. Furthermore, the proposed alternating algorithm outperforms the scheme in \cite{zhang2021reconfigurable}, where our algorithm with $b=1$ gets a similar performance as the scheme in \cite{zhang2021reconfigurable} with $b=2$. With the increase of $N$, the goodput is also improved as seen in Fig. \ref{Disphase}.
 
\begin{figure}[t]
  \centering
  \includegraphics[width=0.48\textwidth]{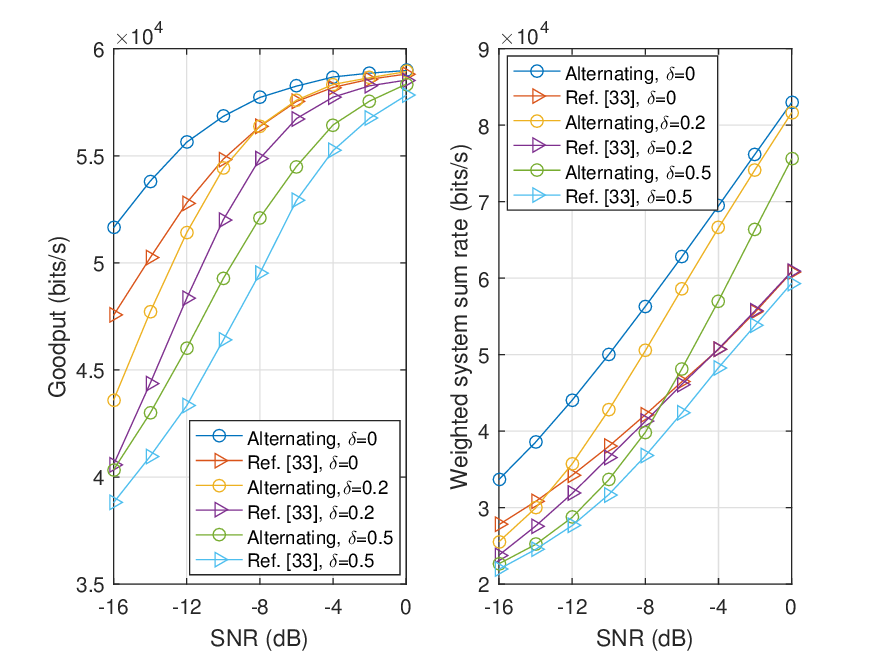}
  \caption{Weighted system sum rate of the alternating algorithm and the scheme in [33] with $K=8$ for (a) $\lambda_0=1$, (b) $\lambda_0=0.5$ under imperfect CSI.} 
  \label{ICSI}
\end{figure}

 Considering the more practical imperfect CSI case, we present the goodput and the weighted system sum rate of the proposed alternating algorithm under imperfect CSI in Fig. \ref{ICSI}, and make the comparison with  the perfect CSI case. The channel estimation errors can be modeled by $\mathbf{H}_{e}=\mathbf{H}_{per}+\mathbf{H}_{m}$, where $\mathbf{H}_{per}$ is the real channel matrix and $\mathbf{H}_{m}\sim \mathcal{CN}(0,\delta^2 N_0 \mathbf{I}_m)$ is channel error matrix\cite{Guo2016,Guo2020}. It indicates that the channel estimation errors variance is propositional to the channel noise variance. We can see that all systems are affected by channel estimation errors. 
In Fig. \ref{ICSI}(a), the proposed algorithm with $\delta=0.2$ outperforms the scheme in \cite{zhang2021reconfigurable} about 2-2.5 dB gains at low SNR region, while for $\delta=0.5$, the performance gains become 2-3 dB. In Fig. \ref{ICSI}(b), the performance of the proposed algorithm with $\delta=0.2$ even outperforms the scheme in \cite{zhang2021reconfigurable} with perfect CSI at high $\mathrm{SNR}$ region.

\section{Conclusions}\label{CON}

A novel RIS-enabled JBAC system has been proposed in this paper, which not only realizes the information-carrying capability at the RIS but also enables BS to receive more desired information. The alternating algorithm is adopted to solve the formulated MOOP and maximize the user average transmission rate of the primary communication system and the goodput of the backscatter communication system simultaneously. With the proposed alternating algorithm, the passive reflecting patterns at the RIS and the active beamforming at the BS are jointly designed. The superiorities of the proposed system in improving the goodput and the user average transmission rate have also been proved through the comparison with the existing RIS based information-carrying systems even under the limited discrete phase shifts and the imperfect CSI. With the increase of the weighting parameter, the goodput of the backscatter communication will be improved. The size of the reflecting candidate set is critical for promoting the goodput of the backscatter communication, while the number of reflecting elements is not. On the contrary, the user average transmission rate of the primary communication can be promoted by increasing the number of reflecting elements rather than the reflecting candidate size. 


\bibliographystyle{IEEEtran} 
\bibliography{IEEEabrv,bib}

\begin{thebibliography}{10}
\providecommand{\url}[1]{#1}
\csname url@samestyle\endcsname
\providecommand{\newblock}{\relax}
\providecommand{\bibinfo}[2]{#2}
\providecommand{\BIBentrySTDinterwordspacing}{\spaceskip=0pt\relax}
\providecommand{\BIBentryALTinterwordstretchfactor}{4}
\providecommand{\BIBentryALTinterwordspacing}{\spaceskip=\fontdimen2\font plus
\BIBentryALTinterwordstretchfactor\fontdimen3\font minus
  \fontdimen4\font\relax}
\providecommand{\BIBforeignlanguage}[2]{{%
\expandafter\ifx\csname l@#1\endcsname\relax
\typeout{** WARNING: IEEEtran.bst: No hyphenation pattern has been}%
\typeout{** loaded for the language `#1'. Using the pattern for}%
\typeout{** the default language instead.}%
\else
\language=\csname l@#1\endcsname
\fi
#2}}
\providecommand{\BIBdecl}{\relax}
\BIBdecl

\bibitem{Basar2020}
E.~Basar, ``Reconfigurable intelligent surface-based index modulation: A new
  beyond {MIMO} paradigm for {6G},'' \emph{{IEEE} Trans. Commun.}, vol.~68,
  no.~5, pp. 3187--3196, Feb. 2020.

\bibitem{Li2022}
G.~Li, L.~Hu, P.~Staat, H.~Elders-Boll, C.~Zenger, C.~Paar, and A.~Hu,
  ``Reconfigurable intelligent surface for physical layer key generation:
  Constructive or destructive?'' \emph{{IEEE} Wireless Commun.}, vol.~29,
  no.~4, pp. 146--153, Aug. 2022.

\bibitem{Basharat2022}
S.~Basharat, S.~A. Hassan, A.~Mahmood, Z.~Ding, and M.~Gidlund,
  ``Reconfigurable intelligent surface-assisted backscatter communication: {A}
  new frontier for enabling {6G} {IoT} networks,'' \emph{{IEEE} Wireless
  Commun.}, pp. 1--8, Jun. 2022.

\bibitem{niu2019overview}
J.-P. Niu and G.~Y. Li, ``An overview on backscatter communications,'' \emph{J.
  Commun. Inf. Netw.}, vol.~4, no.~2, pp. 1--14, Jun. 2019.

\bibitem{Zhang2020a}
Q.~Zhang, Y.-C. Liang, and H.~V. Poor, ``Intelligent user association for
  symbiotic radio networks using deep reinforcement learning,'' \emph{{IEEE}
  Trans. Wireless Commun.}, vol.~19, no.~7, pp. 4535--4548, Jul. 2020.

\bibitem{Liang2022}
Y.-C. Liang, Q.~Zhang, J.~Wang, R.~Long, H.~Zhou, and G.~Yang, ``Backscatter
  communication assisted by reconfigurable intelligent surfaces,'' \emph{Proc.
  IEEE}, pp. 1--19, May 2022.

\bibitem{belo2019iq}
D.~Belo, R.~Correia, Y.~Ding, S.~N. Daskalakis, G.~Goussetis, A.~Georgiadis,
  and N.~B. Carvalho, ``{IQ} impedance modulator front-end for low-power
  {L}o{R}a backscattering devices,'' \emph{IEEE Trans. Microw. Theory Techn.},
  vol.~67, no.~12, pp. 5307--5314, Dec. 2019.

\bibitem{daskalakis2018four}
S.~N. Daskalakis, R.~Correia, G.~Goussetis, M.~M. Tentzeris, N.~B. Carvalho,
  and A.~Georgiadis, ``Four-{PAM} modulation of ambient {FM} backscattering for
  spectrally efficient low-power applications,'' \emph{IEEE Trans. Microw.
  Theory Techn.}, vol.~66, no.~12, pp. 5909--5921, Dec. 2018.

\bibitem{Xu2023}
Y.~Xu, R.~Xu, D.~Li, G.~Yang, G.~Wang, C.~Yuen, and J.~Zhou, ``Robust resource
  allocation for wireless-powered backscatter communication systems with
  {NOMA},'' \emph{{IEEE} Trans. Veh. Technol.}, pp. 1--13, Apr. 2023.

\bibitem{ElMossallamy2019}
M.~A. ElMossallamy, M.~Pan, R.~Jäntti, K.~G. Seddik, G.~Y. Li, and Z.~Han,
  ``Noncoherent backscatter communications over ambient {OFDM} signals,''
  \emph{{IEEE} Trans. Commun.}, vol.~67, no.~5, pp. 3597--3611, May 2019.

\bibitem{Xu2021a}
Y.~Xu, B.~Gu, and D.~Li, ``Robust energy-efficient optimization for secure
  wireless-powered backscatter communications with a non-linear {EH} model,''
  \emph{{IEEE} Commun. Lett.}, vol.~25, no.~10, pp. 3209--3213, Oct. 2021.

\bibitem{Xu2021}
Y.~Xu, Z.~Qin, G.~Gui, H.~Gacanin, H.~Sari, and F.~Adachi, ``Energy efficiency
  maximization in {NOMA} enabled backscatter communications with {QoS}
  guarantee,'' \emph{{IEEE} Wireless Commun. Lett.}, vol.~10, no.~2, pp.
  353--357, Feb. 2021.

\bibitem{Wu2020}
Q.~Wu and R.~Zhang, ``Towards smart and reconfigurable environment: Intelligent
  reflecting surface aided wireless network,'' \emph{{IEEE} Commun. Mag.},
  vol.~58, no.~1, pp. 106--112, Feb. 2020.

\bibitem{xurisa}
Y.~Xu, G.~Gui, H.~Gacanin, and F.~Adachi, ``A survey on resource allocation for
  {5G} heterogeneous networks: Current research, future trends, and
  challenges,'' \emph{IEEE Commun. Surv. Tutor.}, vol.~23, no.~2, pp. 668--695,
  Feb. 2021.

\bibitem{Huang2020}
C.~Huang, S.~Hu, G.~C. Alexandropoulos, A.~Zappone, C.~Yuen, R.~Zhang, M.~D.
  Renzo, and M.~Debbah, ``Holographic {MIMO} surfaces for {6G} wireless
  networks: Opportunities, challenges, and trends,'' \emph{{IEEE} Wireless
  Commun.}, vol.~27, no.~5, pp. 118--125, Jul. 2020.

\bibitem{xurisc}
Y.~Xu, H.~Xie, Q.~Wu, C.~Huang, and C.~Yuen, ``Robust max-min energy efficiency
  for {RIS}-aided hetnets with distortion noises,'' \emph{IEEE Trans. on
  Commun.}, vol.~70, no.~2, pp. 1457--1471, 2022.

\bibitem{xurisb}
Y.~Xu, Z.~Gao, Z.~Wang, C.~Huang, Z.~Yang, and C.~Yuen, ``{RIS}-enhanced
  {WPCNs}: Joint radio resource allocation and passive beamforming
  optimization,'' \emph{{IEEE} Trans. Veh. Technol.}, vol.~70, no.~8, pp.
  7980--7991, Jul. 2021.

\bibitem{Guo2023}
K.~Guo, M.~Wu, X.~Li, H.~Song, and N.~Kumar, ``Deep reinforcement learning and
  {NOMA}-based multi-objective {RIS}-assisted {IS-UAV-TNs}: Trajectory
  optimization and beamforming design,'' \emph{{IEEE} Trans. Intell. Transp.
  Syst.}, pp. 1--14, Apr. 2023.

\bibitem{Peng2022}
Z.~Peng, Z.~Zhang, L.~Kong, C.~Pan, L.~Li, and J.~Wang, ``Deep reinforcement
  learning for {RIS}-aided multiuser full-duplex secure communications with
  hardware impairments,'' \emph{{IEEE} Internet Things J.}, vol.~9, no.~21, pp.
  21\,121--21\,135, May 2022.

\bibitem{Tong2021}
X.~Tong, Z.~Zhang, J.~Wang, C.~Huang, and M.~Debbah, ``Joint multi-user
  communication and sensing exploiting both signal and environment sparsity,''
  \emph{{IEEE} J. Sel. Topics Signal Process.}, vol.~15, no.~6, pp. 1409--1422,
  Nov. 2021.

\bibitem{Liu2022}
Q.~Liu, M.~Fu, W.~Li, J.~Xie, and M.~Kadoch, ``{RIS}-assisted ambient
  backscatter communication for sagin {IoT},'' \emph{{IEEE} Internet Things
  J.}, vol.~10, no.~11, pp. 9375--9384, Nov. 2023.

\bibitem{Fara2022}
R.~Fara, P.~Ratajczak, D.-T. Phan-Huy, A.~Ourir, M.~Di~Renzo, and J.~de~Rosny,
  ``A prototype of reconfigurable intelligent surface with continuous control
  of the reflection phase,'' \emph{{IEEE} Wireless Commun.}, vol.~29, no.~1,
  pp. 70--77, Feb. 2022.

\bibitem{Asiedu2022}
D.~K.~P. Asiedu and J.-H. Yun, ``Multiuser {NOMA} with multiple reconfigurable
  intelligent surfaces for backscatter communication in a symbiotic cognitive
  radio network,'' \emph{{IEEE} Trans. Veh. Technol.}, pp. 1--16, Dec. 2022.

\bibitem{zuo2021reconfigurable}
J.~Zuo, Y.~Liu, L.~Yang, L.~Song, and Y.-C. Liang, ``Reconfigurable intelligent
  surface enhanced {NOMA} assisted backscatter communication system,''
  \emph{IEEE Trans. Veh. Technol.}, vol.~70, no.~7, pp. 7261--7266, Jul. 2021.

\bibitem{chen2021joint}
H.~Chen, G.~Yang, and Y.-C. Liang, ``Joint active and passive beamforming for
  reconfigurable intelligent surface enhanced symbiotic radio system,''
  \emph{IEEE Wireless Commun. Lett.}, vol.~10, no.~5, pp. 1056--1060, May 2021.

\bibitem{jia2020intelligent}
X.~Jia, J.~Zhao, X.~Zhou, and D.~Niyato, ``Intelligent reflecting surface-aided
  backscatter communications,'' in \emph{Proc. IEEE Globecom}, Dec. 2020, pp.
  1--6.

\bibitem{jia2021irs}
X.~Jia and X.~Zhou, ``{IRS}-assisted ambient backscatter communications
  utilizing deep reinforcement learning,'' \emph{IEEE Wireless Commun. Lett.},
  vol.~10, no.~11, pp. 2374--2378, Nov. 2021.

\bibitem{jia2021intelligent}
X.~Jia, X.~Zhou, D.~Niyato, and J.~Zhao, ``Intelligent reflecting
  surface-assisted bistatic backscatter networks: Joint beamforming and
  reflection design,'' \emph{IEEE Trans. Green Commun. and Netw.}, vol.~6,
  no.~2, pp. 799--814, Jun. 2022.

\bibitem{chen2021performance}
Y.~Chen, ``Performance of ambient backscatter systems using reconfigurable
  intelligent surface,'' \emph{IEEE Commun. Lett.}, vol.~25, no.~8, pp.
  2536--2539, Aug. 2021.

\bibitem{nemati2020short}
M.~Nemati, J.~Ding, and J.~Choi, ``Short-range ambient backscatter
  communication using reconfigurable intelligent surfaces,'' in \emph{Proc.
  IEEE WCNC}, May 2020, pp. 1--6.

\bibitem{zhou2020cooperative}
H.~Zhou, Y.-C. Liang, X.~Kang, and S.~Sun, ``Cooperative beamforming for large
  intelligent surface assisted symbiotic radios,'' in \emph{Proc. IEEE
  Globecom}, Dec. 2020, pp. 1--6.

\bibitem{hu2020reconfigurable}
J.~Hu, Y.-C. Liang, and Y.~Pei, ``Reconfigurable intelligent surface enhanced
  multi-user {MISO} symbiotic radio system,'' \emph{IEEE Trans. Commun.},
  vol.~69, no.~4, pp. 2359--2371, Apr. 2021.

\bibitem{zhang2021reconfigurable}
Q.~Zhang, Y.-C. Liang, and H.~V. Poor, ``Reconfigurable intelligent surface
  assisted {MIMO} symbiotic radio networks,'' \emph{IEEE Trans. on Commun.},
  vol.~69, no.~7, pp. 4832--4846, Jul. 2021.

\bibitem{park2020intelligent}
S.~Y. Park and D.~In~Kim, ``Intelligent reflecting surface-aided phase-shift
  backscatter communication,'' in \emph{Proc. IEEE IMCOM}, Jan. 2020, pp. 1--5.

\bibitem{Guo2020}
S.~Guo, S.~Lv, H.~Zhang, J.~Ye, and P.~Zhang, ``Reflecting modulation,''
  \emph{{IEEE} J. Sel. Areas Commun.}, vol.~38, no.~11, pp. 2548--2561, Nov.
  2020.

\bibitem{Ye2022}
J.~Ye, S.~Guo, S.~Dang, B.~Shihada, and M.-S. Alouini, ``On the capacity of
  reconfigurable intelligent surface assisted {MIMO} symbiotic
  communications,'' \emph{{IEEE} Trans. Wireless Commun.}, vol.~21, no.~3, pp.
  1943--1959, Mar. 2022.

\bibitem{Wang2008}
H.~Wang, H.~P. Schwefel, and T.~S. Toftegaard, ``History-based adaptive
  modulation for a downlink multicast channel in {OFDMA} systems,'' in
  \emph{Proc. IEEE WCNC}, Apr. 2008, pp. 1588--1592.

\bibitem{DaiL2021}
C.~Hu, L.~Dai, S.~Han, and X.~Wang, ``Two-timescale channel estimation for
  reconfigurable intelligent surface aided wireless communications,''
  \emph{IEEE Trans. on Commun.}, vol.~69, no.~11, pp. 7736--7747, Nov. 2021.

\bibitem{He2020}
Z.-Q. He and X.~Yuan, ``Cascaded channel estimation for large intelligent
  metasurface assisted massive {MIMO},'' \emph{{IEEE} Wireless Commun. Lett.},
  vol.~9, no.~2, pp. 210--214, Oct. 2020.

\bibitem{jing20}
J.~Wang, C.~Jiang, H.~Zhang, Y.~Ren, K.-C. Chen, and L.~Hanzo, ``Thirty years
  of machine learning: The road to pareto-optimal wireless networks,''
  \emph{IEEE Commun. Surv. Tutor.}, vol.~22, no.~3, pp. 1472--1514, Jan. 2020.

\bibitem{zeger1990pseudo}
K.~Zeger and A.~Gersho, ``Pseudo-{G}ray coding,'' \emph{IEEE Trans. commun.},
  vol.~38, no.~12, pp. 2147--2158, Dec. 1990.

\bibitem{Guo2016}
S.~Guo, H.~Zhang, P.~Zhang, and D.~Yuan, ``Link-adaptive mapper designs for
  space-shift-keying-modulated {MIMO} systems,'' \emph{{IEEE} Trans. Veh.
  Technol.}, vol.~65, no.~10, pp. 8087--8100, Dec. 2016.

\bibitem{Luo2023}
H.~Luo, R.~Liu, M.~Li, and Q.~Liu, ``{RIS}-aided integrated sensing and
  communication: Joint beamforming and reflection design,'' \emph{{IEEE} Trans.
  Veh. Technol.}, pp. 1--5, Feb. 2023.

\bibitem{Liu2021}
R.~Liu, M.~Li, Q.~Liu, and A.~L. Swindlehurst, ``Dual-functional
  radar-communication waveform design: A symbol-level precoding approach,''
  \emph{{IEEE} J. Sel. Topics Signal Process.}, vol.~15, no.~6, pp. 1316--1331,
  Sep. 2021.

\end{thebibliography}

\end{document}